\documentclass[useAMS,usenatbib,aas_macros]{mn2e}
\usepackage{aas_macros}
\usepackage{amsmath}

\usepackage{array}
\usepackage{multirow}

\PassOptionsToPackage{normalem}{ulem}
\usepackage{ulem}

\usepackage{color}

\newcommand{\equ}[1]{eq.~(\ref{eq:#1})}
\newcommand{\equs}[1]{eqs.~(\ref{eq:#1})}

\newcommand{\se}[1]{\S\ref{sec:#1}}
\newcommand{\fig}[1]{Fig.~\ref{fig:#1}}
\newcommand{\figs}[1]{Figs.~\ref{fig:#1}}
\newcommand{\Fig}[1]{Figure~\ref{fig:#1}}
\newcommand{\Figs}[1]{Figures~\ref{fig:#1}}
\newcommand{\be}{\begin{equation}}
\newcommand{\ee}{\end{equation}}
\newcommand{\ba}{\begin{align}}
\newcommand{\ea}{\end{align}}
\newcommand{\bad}{\begin{equation} \begin{aligned}}
\newcommand{\ead}{\end{aligned} \end{equation}}
\newcommand{\bea}{\begin{eqnarray}}
\newcommand{\eea}{\end{eqnarray}}

\def\la{\langle}
\newcommand{\tab}[1]{Table~\ref{tab:#1}}

\newcommand{\msun}{M_\odot}
\newcommand{\Msun}{M_\odot}

\newcommand{\ifm}[1]{\relax\ifmmode#1\else$\mathsurround=0pt #1$\fi}
\newcommand{\kms}{\ifmmode\,{\rm km}\,{\rm s}^{-1}\else km$\,$s$^{-1}$\fi}
\newcommand{\hmpc}{\,\ifm{h^{-1}}{\rm Mpc}}

\newcommand{\Myr}{\,{\rm Myr}}

\newcommand{\erg}{\,{\rm erg}}

\newcommand{\ltsima}{$\; \buildrel < \over \sim \;$}
\newcommand{\lsim}{\lower.5ex\hbox{\ltsima}}
\newcommand{\gtsima}{$\; \buildrel > \over \sim \;$}
\newcommand{\gsim}{\lower.5ex\hbox{\gtsima}}
\newcommand{\prop}{\propto}
\newcommand{\dd}{\rm d}

\def\omm{\Omega_{\rm m}}
\def\oml{\Omega_{\Lambda}}
\def\omb{\Omega_{\rm b}}

\def\Mv{M_{\rm v}}

\def\Rv{R_{\rm v}}
\def\Vv{V_{\rm v}}

\def\Fv{F_{\rm v}}

\def\Ms{M_{\rm s}}

\def\rn{\rho_{\rm c}}
\def\rb{\bar\rho_{\rm c}}
\def\rba{\bar\rho_{\rm c,1}}
\def\rbb{\bar\rho_{\rm c,2}}
\def\rbi{\bar\rho_{\rm c,i}}
\def\rs{r_{\rm c}}
\def\brho{\bar\rho}
\def\brhov{\bar\rho_{\rm v}}

\def\tgamma{\bar\gamma}
\def\bs{\bar{s}}
\def\bsi{\bar{s}_1}
\def\bsv{\bar{s}_2}
\def\cm{c_{\rm m}}

\def\Mt{R_{\rm t}}
\def\Rt{R_{\rm t}}
\def\Vt{V_{\rm t}}
\def\ct{c_{\rm t}}
\def\bmu{\mu}

\title[Analytic Halo Profile]
{Dark-Matter Halo Profiles of a General Cusp/Core with Analytic 
Velocity and Potential}

\author[A. Dekel]
{Avishai Dekel$^1$,
Guy Ishai$^1$,
Aaron A. Dutton$^2$,
Andrea V. Macci{\`o}$^{2,3}$\\
\\
$^1$Racah Institute of Physics, The Hebrew University, 
Jerusalem 91904 Israel\\
$^2$New York University Abu Dhabi, PO Box 129188, Abu Dhabi, United Arab
Emirates\\
$^3$Max-Planck-Institut f\"ur Astronomie, K\"onigstuhl 17, 69117 Heidelberg,
Germany\\
dekel@huji.ac.il
}

\begin{document}

\large

\pagerange{\pageref{firstpage}--\pageref{lastpage}} \pubyear{2002}

\maketitle

\label{firstpage}

\begin{abstract}
We present useful functions for the profiles of dark-matter
(DM) haloes with a free inner slope, from cusps to cores, 
where the profiles of density, mass-velocity and {\it potential} are 
simple analytic expressions. 
Analytic velocity is obtained by expressing the mean density 
as a simple functional form, and deriving the local density by 
differentiation. 
The function involves four shape parameters,
with only two or three free:
a concentration parameter $c$,
inner and outer asymptotic slopes $\alpha$ and $\tgamma$,
and a middle shape parameter $\beta$.
Analytic expressions for the potential and velocity dispersion 
exist for $\tgamma=3$ and 
for 
$\beta$ a natural number.
 
We match the models to the DM haloes in cosmological 
simulations, with and without baryons, ranging   
from steep cusps to flat cores. 
Excellent fits are obtained with three free parameters ($c$, $\alpha$,
$\tgamma$) and $\beta = 2$.
For an analytic potential, similar fits 
are obtained for $\tgamma=3$ and $\beta=2$ with only two 
free parameters ($c$, $\alpha$); this is our favorite model.
A linear combination of two such profiles, 
with an additional free concentration parameter,
provides excellent fits also for $\beta=1$, where the expressions are
simpler.
The fit quality is comparable to non-analytic popular models.

An analytic potential is useful for modeling the inner-halo evolution 
due to gas inflows and outflows, studying environmental
effects on the outer halo, and generating halo potentials
or initial conditions for simulations.
The analytic velocity can quantify simulated 
and observed rotation curves without numerical integrations.
\end{abstract}


\section{Introduction}

The shapes of the density profiles of dark-matter haloes, 
as deduced from %
cosmological N-body simulations of DM only (DMO),
are commonly fit by a function with one free parameter, such as
the NFW profile \citep{nfw97}.
This profile has a fixed power-law cusp at small radii with an asymptotic
log slope $-\alpha$ where $\alpha=1$, and a fixed asymptotic slope $-\gamma$ 
with $\gamma=3$ at large radii.
The free parameter is the characteristic radius $\rs$ of the inner cusp/core,
which can be replaced by the concentration parameter $c$, defined by
$c=\Rv/\rs$, where $\Rv$ is the halo virial radius, 
determined by the halo mass and cosmological time.
However, simulated halo profiles, 
especially when baryons are included, %
may show deviations from this universal shape
both in the inner cusp and in the outskirts near the halo virial radius $\Rv$,
as well as in between. %
In particular, the observed halo profiles, especially in low-mass galaxies,
tend to have a flatter cusp, with $\alpha<1$, and possibly even a
constant-density core, $\alpha \simeq 0$
\citep{deblok01,swaters03,goerdt06,walker11,oh11},
while massive galaxies may show a steeper cusp, $\alpha>1$. 
Within the standard cosmology with non-interacting cold dark matter,
the common wisdom is that the baryonic processes associated with galaxy
formation and evolution are responsible for strong evolution of the DM
inner profiles, steepening the cusp in massive haloes and flattening it in
lower-mass haloes, potentially all the way to a flat core with $\alpha=0$
\citep[e.g., simulations by][and references therein]{tollet16}.  
On the other side, environmental tidal effects may alter the halo profile 
in the outskirts \citep[e.g.][]{more15}. 
For the modeling of the halo profile as it evolves between cusp and core, 
or as it is stripped from the outside,
and for quantifying the variety of simulated and observed velocity profiles, 
one desires to have a function with more freedom than the NFW profile. 
In addition to the free concentration parameter,
a free inner slope $\alpha$ can help matching the variations between a cusp 
and a core, 
a free outer slope $\gamma$ may provide a flexibility at the outskirts when 
necessary,
and an intermediate shape parameter $\beta$ may improve the fit in the 
middle halo when very high accuracy is desired.

\smallskip
While the profile has to provide a good fit to the variety of DM-halo profiles
with a minimum number of free parameters, %
our desire is to have {\it analytic} expressions both for the density profile 
and for the integrated mass profile, which immediately translates to
the DM circular velocity profile that can be deduced from observations. 
Furthermore, we wish to have an
analytic expression for the gravitational potential profile,
being crucial, e.g., for the analytic modeling of the evolution between a 
cusp and a core. 
In addition, 
an analytic expression for the velocity-dispersion profile may help
constructing DM haloes in equilibrium, e.g., as initial conditions for
simulations.

\smallskip
Several density profiles with different levels of flexibility in the inner 
slope have been proposed and some are widely used 
\citep{einasto65,jaffe83,hernquist90,dehnen93,tremaine94,evans94,
burkert95,zhao96,jing+suto00,navarro04,stoehr06,merritt06,
dicintio14a,schaller15,oldham16}.   
In particular, the Einasto profile 
\citep[][see \equ{einasto}]{einasto65,navarro04},
with one additional free shape parameter,
provides an excellent fit to the cusps of DM-only simulated haloes.
Unfortunately, it does
not have sufficient flexibility to accommodate inner cores (\se{comparison}).
Among the profiles with a flexible inner slope, 
the profile proposed by \citet{dehnen93} and \citet{tremaine94} 
stands out as having analytic expressions for the density, mass and potential. 
It is useful in modeling spherical stellar systems, but as is,
with only one free shape parameter, it does not have the flexibility for 
a good fit to DM haloes.
This analytic profile can be partly generalized with an additional parameter 
$\beta$
that characterizes the transition region between the asymptotic regions 
\citep{zhao96}, but this by itself still
does not provide the flexibility required for fitting DM haloes.
To the best of our knowledge, the profiles used so far that can resemble the 
variety of DM haloes do not have analytic expressions for the mass-velocity or  
potential profiles.  Analytic expressions are limited  
to special cases, such as the NFW cusp with $\alpha=1$, or a similar profile
with $\alpha=0$, as well as other special cases \citep[summarized
in][]{zhao96}. The desired analytic expressions are missing
for profiles that fit DM haloes with sufficient flexibility
in the inner and outer regions.

\smallskip
Here we propose profiles of the desired analytic nature, 
which fit very well the profiles of DM haloes with a general cusp-core. 
The first has simple analytic expressions for 
the density and mass-velocity profiles in the general case of a free outer
slope. 
The two others also have analytic potentials for a general inner slope $\alpha$ 
with the outer asymptotic slope of the local density $\rho(r)$ 
fixed at either $\gamma=4$ or $\gamma=3.5$, both corresponding to 
a slope $\tgamma=3$ for the mean-density profile $\brho(r)$. 
The asymptotic outer slope,
which is materialized well outside the halo virial radius, is compensated for
by a proper choice of another parameter, the concentration parameter $c$.
The case $\gamma=3.5$, with a proper choice of middle shape parameter
$\beta=2$ (see below), 
provides excellent fits to simulated haloes with only two free parameters.
The case $\gamma=4$ provides adequate fits, which become excellent once
a sum of two such functions is considered, with an additional free
concentration parameter.
Our proposed profiles are inspired by earlier ideas concerning
analytic integrals \citep[e.g][]{dehnen93,tremaine94,zhao96}, 
combined with a concentration parameter \citep[e.g.][]{nfw97}, 
and if necessary a linear combination of two functions
\citep[e.g.][]{zhao96,schaller15}.

\smallskip
In \se{flexible} we present the flexible profile with analytic density and
mass-velocity.
In \se{tgamma3} we present the profiles with fixed outer slopes that
also provide analytic expressions for the potential and velocity dispersion 
profiles, with either two or three free parameters.
In \se{sims} we compare the fits of the different proposed model profiles to
simulated halo profiles with and without baryons, spanning a variety of cusps 
and cores. We then compare the new analytic models to
other fitting functions that do not have analytic expressions for
mass-velocity and potential.
In \se{conc} we summarize and discuss our results.






\section{A Flexible Profile with Analytic Mass-Velocity and Density}
\label{sec:flexible}

\subsection{Introduction: A General Non-analytic Profile}
\label{sec:flex_intro}

As an introductory reference,
consider the very flexible and commonly used functional form for the shape of 
the density profile (sometimes termed the $\alpha\beta\gamma$\footnote{Note 
that different authors may use different permutations of these parameters.}
profile),
\be
\rho(r)=\frac{\rn}{x^\alpha(1+x^{1/\beta})^{\beta(\gamma-\alpha)}}, 
\label{eq:flex}
\ee
where we scale the radius by an intermediate 
radius $\rs$, related to $\Rv$ by a concentration parameter $c$,
\be
x=\frac{r}{\rs}, \quad \rs=\frac{\Rv}{c}.
\ee
The DM-halo virial radius $\Rv$ is the physical scale determined by cosmology
at a given time for a given halo mass, so for the sake of studying halo
profile shapes we measure distances $r$ with respect $\Rv$, and replace $\rs$
by $c$ as a free parameter.
The parameters $\alpha$ and $\gamma$ are the asymptotic slopes 
of $\log\rho(r)$ at $x\ll 1$ and $x\gg 1$ respectively.
The parameter $\beta$ characterizes the shape near the transition radius
$x \sim 1$.
The characteristic density $\rn$ can be 
expressed in terms of $\brhov$, the mean mass density within $\Rv$, 
defined to be, e.g., a factor of 200 larger than the cosmological critical 
mean density.
This functional form thus has in principle four shape parameters, 
$\alpha$, $\beta$, $\gamma$ and $c$.
We will see in \se{sims} that for the purpose of fitting DM haloes 
from simulations with analytic expressions one can do with three and even 
only two free parameters.
As will be discussed in \se{physical},
the parameters in \equ{flex} do not necessarily have a straightforward physical
meaning, but they can be replaced by more physical parameters.

\smallskip
The general functional form of \equ{flex} reduces to the standard NFW profile
\citep{nfw97}
for $\alpha=1$, $\beta=1$, and $\gamma=3$,
leaving $c$ as the single free shape parameter.
In this case the slope of $\rho(r)$ at $\rs$ is $-2$.
The NFW profile has been extremely useful in fitting the density 
profiles of DM haloes in cosmological simulations of dark matter 
only (DMO) with no baryons.
With one or more of the additional parameters free, 
\equ{flex} may provide the flexibility required for fitting the profiles 
of haloes that have been modified by baryonic processes or environmental 
effects.

\smallskip
The associated profiles of mass and velocity are needed, e.g., for
comparison with observed rotation curves.  The associated profiles of
potential and velocity dispersion are needed, e.g., for analytic modeling of 
halo evolution.
In special cases, e.g., when 
$\alpha$, $\beta$ and $\gamma$ are natural numbers, as in NFW, 
it may be possible to obtain analytic expressions for all these profiles.
However, observed haloes, as well as haloes simulated with baryons or in
clustered environments, require that the parameters, especially $\alpha$,
are general real numbers.
In most such cases, one has to perform numerical integrations of \equ{flex}  
in order to yield the mass-velocity, potential and velocity dispersion
profiles. 
The same is true for most other functional forms that have been used to fit DM
haloes.
This includes in particular the Einasto profile 
\citep{einasto65,navarro04,merritt06,graham06,gao08,dutton14},
which provides better fits than NFW to DMO-simulated profiles.
It also includes other profiles that allow a match to simulated profiles
which deviate from the NFW or the Einasto profiles
\citep[e.g.][]{dicintio14a,schaller15}. 

\smallskip
In order to enable straightforward comparisons to observed rotation curves,
and in order to quantify the effects of baryons on the inner halo and 
environment on the outer halo,
we seek a functional form with free inner and outer slopes in which the 
profiles of density and mass-velocity are given by analytic expressions. 
For the purpose of an analytic study of the evolution of the inner halo due 
to baryonic processes, we require that the potential profile 
should 
also be analytic.
An analytic isotropic velocity dispersion will enable constructing a DM halo in
equilibrium.
In the following subsection we propose a modification of the flexible
multi-parameter profile of \equ{flex} that allows analytic expressions for the
mass and velocity profiles.
Then in \se{tgamma3} we introduce special cases of this profile which also
have analytic potential and velocity dispersion and are very useful in
fitting the variety of DM halo profiles with only two or three free shape 
parameters.

\subsection{Analytic Mass-Velocity and Density}
\label{sec:flex_brho}

\subsubsection{Mean density}

The idea for obtaining analytic density and mass-velocity profiles 
is very simple.  
We apply a functional form inspired by \equ{flex}, but to the
{\it mean} density within the sphere of radius $r$ (or equivalently 
to the mass or velocity profiles) rather than to the local density at $r$, 
namely
\be
\brho(r) = \frac{\rb}{x^{\alpha} (1+x^{1/\beta})^{\beta(\tgamma-\alpha)}} ,
\quad x=\frac{r}{\rs}, \quad \rs=\frac{\Rv}{c} .
\label{eq:brho}
\ee
The local density profile is then determined by a straightforward derivative.
The mass, velocity and force profiles are derived straightforwardly from
$\brho(r)$. For certain specific choices of $\beta$ and $\tgamma$,
the potential profile is derived by analytic integration, for a general value
of $\alpha$ (\se{tgamma3}). 

\smallskip
Now the parameters refer to $\brho(r)$ rather than $\rho(r)$,
and we explicitly distinguish the asymptotic outer slopes $\tgamma$ and 
$\gamma$ in \equ{brho} and \equ{flex}.
The parameter $c$, as in \equ{flex}, refers to an inner radius, 
$\rs=\Rv/c$, that marks the middle-halo transition between the asymptotic 
slopes of $\brho(r)$, 
though the slope there, for either $\brho$ or $\rho$, is in general not $-2$
(as it is for NFW), so its interpretation as a characteristic intermediate
radius can be dubious. 
For the purpose of studying the profile shape, we measure the radius $r$
in terms of the virial radius $\Rv$, so that $\rs$ in \equ{brho}
is replaced by $c^{-1}$.

\smallskip
The normalization factor $\rb$ in \equ{brho} is expressed as a function 
of $\brhov$ and the shape parameters, 
\be
\rb = c^3 \bmu \brhov,  
\label{eq:rb}
\ee
where 
\be
\mu = \frac{(1+c^{1/\beta})^{\beta(\tgamma-\alpha)}} {c^{(3-\alpha)}} .
\label{eq:mu}
\ee
For the purpose of comparing profile shapes, 
we measure the density by means of $\brhov$, 
and the mass and velocity by means of
the virial mass $\Mv$ and velocity $\Vv$.\footnote{Recall that at a given 
cosmological time, for a given $\Delta_{\rm v}=200$ defining the
virial radius, there is a one-to-one correspondence between all the virial
quantities.}
The shape of this very flexible profile can thus involve four parameters:
$\alpha$, $\beta$, $\tgamma$ and $c$.

\smallskip
If the profile has a power-law inner cusp or core, the same $\alpha$ represents 
the asymptotic inner slope of both $\brho$ and $\rho$. 
In the outer asymptote,
we first address here the profiles of density, mass and velocity for a general
$\tgamma$, which are very flexible in matching the outskirts of haloes
subject to environmental effects.
Then in \se{tgamma3} we appeal to special cases of $\tgamma=3$, which have in
addition analytic potentials and still provide excellent fits to the variety of
DM halo profiles with only two or three free parameters.

\subsubsection{Mass, Velocity and Force}

The mass profile is easily deduced from \equ{brho},
\be
M(r) = \frac{4\pi}{3} \brho(r) r^3 = \mu \Mv\, x^3 \brho(r)/\rb .
\label{eq:M}
\ee
The velocity profile, the common observable,
immediately follows (adopting hereafter $G=1$),
\be
V^2(r) = \frac{M(r)}{r} =
c \mu \Vv^2\, x^2 \brho(r)/\rb ,
\label{eq:V}
\ee
where $\Vv^2=\Mv/\Rv$.
The force profile is
\be
F(r) = -\frac{M(r)}{r^2} =
c^2 \mu \Fv\, x \brho(r)/\rb ,    
\label{eq:F}
\ee
where $\Fv=-\Mv/\Rv^2$.
Note that the maximum velocity is obtained at
\be
x_{\rm max} = \left( \frac{2-\alpha}{\tgamma-2} \right) ^{\beta} ,
\ee
which is also where the slope of $\brho(r)$ is $-2$, but generally not 
where the slope of the local $\rho(r)$ is $-2$ (see \se{slopes} below).
For $\alpha+\gamma=4$ the peak velocity and mean-density slope of $-2$
coincide at $x=1$.
The maximum value of the velocity, for $c\gg 1$, 
is $V_{\rm max}^2 \propto c^{\tgamma-2}$.

\subsubsection{Local Density}

\smallskip
The local density profile is obtained from the mass profile by derivative,
\be
\rho(r) = \frac{1}{4\pi r^2} \frac{dM}{dr},
\ee
namely,
\be
\begin{aligned}
\rho(r)=\frac{3-\alpha}{3} 
\left( 1+\frac{3-\tgamma}{3-\alpha}\, x^{1/\beta} \right)
\frac{1}{(1+x^{1/\beta})}\, \brho(r) .
\label{eq:rho}
\end{aligned}
\ee

\smallskip
For a general $\tgamma \neq 3$ this does not resemble the functional form of
\equ{flex} as the term in big parentheses involves a sum of two different 
powers of $x$. 
In the asymptotic inner-halo limit, $x \ll 1$, 
we do have $\rho \prop x^{-\alpha}$, as in $\brho(r)$, with
\be
\rho \simeq \frac{\rn}{x^\alpha}, \quad
\brho \simeq \frac{\rb}{x^\alpha} , \quad
\rn \simeq \frac{3-\alpha}{3} \rb , \quad
V^2 \simeq c\bmu \Vv^2 x^{2-\alpha} .
\ee
\smallskip
In the asymptotic outskirts, $x \gg 1$, once $\tgamma \neq 3$,
we have $\rho(r) \propto \brho(r)$, namely 
$\rho \prop x^{-\gamma}$ with $\gamma=\tgamma$.
However, for $\tgamma$ near 3, this slope may be materialized only well
beyond the virial radius.

\smallskip
For the special case $\tgamma=3$, which allows an analytic potential,
\equ{rho} becomes the same as \equ{flex}, with
\be
\gamma = 3+\beta^{-1} .
\ee
If the asymptotic slope is steeper than the desired slope near and inside
$\Rv$,
it could be partly compensated for by a proper choice of a lower value for $c$.
Otherwise, a more accurate match in the outer
regions may be helped by a deviation of $\tgamma$ from 3.

\subsubsection{Slopes}
\label{sec:slopes}

The parameters $\alpha$ and $\tgamma$ (or $\gamma$) are the slopes in the
asymptotic regions, which may fall well outside the radius range of interest,
for example between $0.01\Rv$ and $\Rv$. For the slopes in points of interest,
the slope profile of $\brho(r)$ is $-\bs (r)$, derived from \equ{brho}
to be 
\be 
\bs (r) = -\frac{\dd \log \brho}{\dd \log r}
= \frac{\alpha+\tgamma\, x^{1/\beta}}{1+x^{1/\beta}} .
\label{eq:bslope}
\ee
This allows one to express the slopes in specific regions of interest
by the model parameters. 
Asymptotically, at $x \ll 1$ the slope corresponds to $\bs=\alpha$
and at $x \gg 1$ it is $\bs=\tgamma$.  
At $x=1$, we have $\bs=0.5(\alpha+\tgamma)$,
reducing to $\bs=2$ (for $\brho$, not $\rho$)
when $\alpha+\tgamma=4$ (e.g. $\alpha=1$ and $\tgamma=3$).
Inverting \equ{bslope},
a slope of $-\bs$ is obtained by $\brho(r)$ at
\be
x_{-\bs} = \left( \frac{\bs-\alpha}{\tgamma-\bs} \right) ^{\beta}  .
\label{eq:bxs}
\ee
In particular, $\bs=2$, where the velocity curve is at a peak, is
obtained at
\be
x_{-\bar2} = x_{\rm max} = \left( \frac{2-\alpha}{\tgamma-2} \right) ^{\beta} .
\label{eq:bx2}
\ee
This defines an alternative and more physical characteristic radius 
$r_{\rm max}$,
which coincides with $\rs$ for $\alpha+\tgamma=4$.
The corresponding alternative concentration parameter is
\be 
\cm = \frac{\Rv}{r_{\rm max}}
= \left( \frac{\tgamma-2}{2-\alpha} \right)^\beta c ,
\label{eq:cmax}
\ee
coinciding with $c$ only for $\alpha+\tgamma=4$.

\medskip
The slope profile of $\rho(r)$ can be similarly derived from \equ{rho}.
For example, for $\tgamma=3$ it is
\be
s(r)= \frac{\alpha+\gamma x^{1/\beta}}{1+x^{1/\beta}} , \quad
\gamma=\tgamma+\beta^{-1} .
\label{eq:slope}
\ee
While at $x \ll 1$ the asymptotic slope is $\alpha$, the same as for $\brho$,
at $x \gg 1$ it is $\gamma$, which is steeper than the $\tgamma=3$, and  
becomes somewhat closer to it for larger $\beta$.
A slope of $s=2$ for $\rho(r)$ is obtained in this case at
\be
x_{-2} = \left( \frac{2-\alpha}{\gamma-2} \right) ^\beta ,
\ee
with $\gamma$ replacing $\tgamma$ in \equ{bx2}.
This is $x=1$ for $\alpha+\gamma=4$.
The slope of $\rho(r)$ deviates from the slope of $\brho(r)$ by 
\be
\Delta s = s(r) - \bs(r) = \frac{\beta^{-1} x^{1/\beta}} {1+x^{1/\beta}} .
\label{eq:Ds}
\ee
At $r=0.015\Rv$,
with $c \sim 10$, typical in the fits to cuspy profiles, 
this deviation is only $\Delta s \sim 0.1$. 
With a larger $c$, and with a larger $\beta$, the deviation could be larger 
by a factor of $\sim 2-3$.

\smallskip
We will see in \se{sims} that \equ{brho}, with a fixed $\beta$ in the range
$1-3$, and especially with $\beta \sim 2$,
provides excellent fits to simulated profiles, where $\alpha$,
$c$ and $\tgamma$ are free. 
The challenge next is to obtain an explicit expression for the potential.
This can be done for a general $\alpha$ with specific choices of $\beta$
and $\tgamma$, as we show in \se{tgamma3}.

\subsubsection{Physical Meaning of the Parameters}
\label{sec:physical}

The values of the parameters in the functional form of \equ{brho},
as obtained by a best fit to a simulated or observed target 
profile within a given radius
range of interest, say $(0.01-1)\Rv$, may not have an obvious physical meaning.
For example, in some cases, the asymptotic slope $\alpha$ may be materialized
only well below the minimum radius of interest, while the quantity of interest 
is the slope near this minimum radius, which can in principle be very different 
from $\alpha$. Similarly, the value of $c$ (namely $\rs$) may be hard to 
interpret, and in some cases $c$ could be so large such that $\rs$ is below 
the minimum radius of interest. 
Quantities of physical meaning are, for example, the slopes at the inner
and outer radii of interest, say $\bsi$ and $\bsv$
at $r=0.015\Rv$ and at $\Rv$, respectively,
as well as the concentration parameter referring to the maximum velocity, 
$\cm$.
These quantities are given as functions of the model parameters in
\equ{bslope} and \equ{cmax}.
Note that a model that matches the target profile within the range of interest 
may in principle deviate from a real DM-halo profile outside this range, 
and in some cases 
be totally irrelevant there. The moral is that the best-fit model
should not be extrapolated without care to outside the radius range within 
which the best fit was performed.

\smallskip
In order to demonstrate the possibly dubious physical meaning of the model 
parameters and how large the deviations outside the fitting range could be,
one can estimate $N$ best-fit parameters from $N$ pairs of radii
and the given slopes at these radii, $(r_{-\bs},\bs)$,
in a target profile from simulations or observations. 
One can apply to each pair \equ{bslope} (or \equ{bxs}), e.g. in the form 
\be
c= \left( \frac{\bs-\alpha}{\tgamma-\bs} \right) ^{\beta} \frac{\Rv}{r_{-\bs}},
\label{eq:cs}
\ee
and solve the set of $N$ equations for the free model parameters, $c$,
$\alpha$, and so on.
As a simple example, we fix in the model $\tgamma=3$ and 
$\beta$ at either $1$ or $2$, 
and solve the corresponding set of two equations for
a target with given $\bsi$ and $\bsv=2.3$ (typical in simulated haloes).

\smallskip
For a cusp of $\bsi=1$, 
the solution for $\beta=1$ is $(\alpha,c)=(0.94,1.9)$, 
namely $\alpha$ is close to the target inner slope but $c$ is rather small,
with $\rs$ not much smaller than $\Rv$.
For a similar cusp but with $\beta=2$ the solution becomes $(0.30,8.2)$, 
namely $\alpha$ is very different from the target slope of $\bsi=1$, 
while the value of 
$c$ is closer to what one expects from $\cm$ or from
concentrations obtained for the NFW profile.

\smallskip
For a core of $\bsi=0$,
the solution for $\beta=1$ is $(\alpha,c)=(-0.16,3.5)$, 
and for $\beta=2$ it is $(-2.5,48)$. In the latter, 
$\alpha$ is very different from $\bsi$ and $c$ is very large, making
$\rs$ not much larger than the minimum radius of interest $0.01\Rv$.
In this case, of fitting a core with $\beta=2$,
the value of $\alpha$ becomes even more negative and $c$ becomes even larger
when the constraint $(\Rv,\bsv)$ is replaced by $(r_{-\bar{2}},2)$
where the velocity peaks, which is typically at $r_{-\bar{2}}/\Rv = 0.16$ 
in simulated haloes.
Indeed, best fits to simulated profiles with cores, in \se{sims},
yield large negative $\alpha$ values and $c \sim 100$ or even larger.
The same is true for $\beta=2$ when $\tgamma$ is left free in the fit.
In this case, of $\beta=2$ fitting a core,
the profile at radii below $0.01\Rv$ is unphysical,
with a mean-density profile that is rising with radius encompassing a hole at 
very small radii.
The virtue of such models with $\beta=2$ is the excellent fit they provide
in the range of interest to the variety of halo profiles, 
and their fully analytic nature when applied with $\tgamma=3$ (see below).

\section{Analytic Potential and Dispersion}
\label{sec:tgamma3}

\subsection{Special Cases with Fully Analytic Solutions} 

The density profile of \equ{flex} has fully analytic expressions for the 
profiles
of mass-velocity, potential and velocity dispersion in the special cases where 
$\beta=n$ and $\gamma=3+k/n$ with $n$ and $k$ natural numbers ($1,2,...$).
The cases with $k=1$ are equivalent to $\tgamma=3$ in \equ{brho} for any $n$.
These expressions for general $n$ and $k$ are provided in detail in 
\citet{zhao96}, and are summarized in our appendix \se{zhao}. 
Originally these profiles
were meant to fit the stellar profiles of spheroidal galaxies, where there was 
no need to scale the radius by a variable $\rs$, or equivalently by 
a free concentration parameter $c$. 
In particular, a good fit is obtained to ``classical" stellar
spheroids by the special case $k=n=1$, where the outer slope is rather steep, 
$\gamma=4$, and the inner slope is free \citep{dehnen93,tremaine94}. 
This case is analogous to \equ{brho} with $\beta=1$ and $\tgamma=3$, except that
we add a free concentration parameter $c$, which also helps dealing with the 
otherwise too steep outer slope in the context of DM haloes.
For the purpose of DM haloes, where the outer slope is typically less steep
than for spheroidal stellar systems, we restrict our attention here to two 
models of this family, both with $\tgamma=3$ ($k=1$), 
one with $n=1$ ($\gamma=4$), and the other 
with $n=2$ ($\gamma=3.5$).
For any $n$, the lowest $k$ guarantees that $\gamma$ is the smallest, 
and for larger values of $n$ $\gamma$ gets smaller and closer to $3$.
However, $n\geq 3$ correspond to shapes that do not match DM halo profiles.

\smallskip
For $k=1$ and a general $n$, the density profile is
\be
\rho(r) = \frac{\rn}{x^{\alpha}\,(1+x^{1/n})^{n(3+1/n-\alpha)}} .
\ee
The mean-density profile is
\be
\brho(r) = \frac{\rb}{x^\alpha \, (1+x^{1/n})^{n(3-\alpha)}} ,
\label{eq:brhon}
\ee 
namely as in \equ{flex} and \equ{brho} with $\beta=n$ and $\tgamma=3$,
and with
$3 \rn=(3-\alpha) \rb$. 
The mass, velocity and force profiles are given in \equs{M} to (\ref{eq:F}),  
with $\rb$ and $\mu$ from \equ{rb} and \equ{mu}, 
substituting $\beta=n$ and $\tgamma=3$.
Recall that the outer asymptotic slope for $\brho$ is $\tgamma=3$
for any $n$, while the asymptotic slope for $\rho$ is $\gamma=3+1/n$.

\vfill\eject
\subsection{Two-parameter Potential for $\tgamma=3$ and $\beta=1$}
\label{sec:beta1}

Here we fix $n=1$, so with $k=1$ the density profiles are simply
\be
\rho(r) = \frac{\rn}{x^\alpha\, (1+x)^{4-\alpha}} , \quad
\brho(r) = \frac{\rb}{x^\alpha\, (1+x)^{3-\alpha}} ,
\ee
With one free parameter, $\alpha$, this resembles elliptical galaxies.
With two free parameters, $\alpha$ and $c$, we will see in \se{sims} that
this function provides reasonable matching to simulated profiles in the inner
DM halo, though with possible $\sim 10\%$ deviations in the middle halo.
The mass, velocity and force profiles are given in \equs{brho} to (\ref{eq:F})
with $\tgamma=3$ and $\beta=2$.

\smallskip
The potential is obtained by integration of the force over radius. 
We assume that the halo density profile is truncated at a certain radius $\Rt$
(which could be $\sim \Rv$ or larger, as desired).
The potential at $r\leq \Rt$, defined to vanish at infinity, is given by
\be
U(r) = -\int_r^{\Rt} \frac{M(y)}{y^2} \dd y 
       - \int_{\Rt}^\infty \frac{\Mt}{y^2} \dd y .
\ee
Denoting 
$\ct= (\Rt/\Rv)\, c$, 
$\Vt=V(\Rt)$ and $\Mt=M(\Rt)$ from \equ{V} and \equ{M}
with $\tgamma=3$ and $\beta=1$,
one obtains
\be
\begin{aligned}
U(r) =& -\Vt^2 + \frac{c\mu}{(2-\alpha)} \Vv^2 \times \\
      &\left[ \left( \frac{x}{1+x} \right) ^{2-\alpha}
             -\left( \frac{\ct}{1+\ct} \right) ^{2-\alpha} \right] ,
\end{aligned}
\ee
where $\mu$ is given by \equ{mu} for $\tgamma=3$ and $\beta=1$.
%
%
The second term in the square brackets ensures that $U(\Rt)=-\Vt^2$,
and the normalization factor in front of the square brackets guarantees that
$\dd U/\dd r = -F(r)$. 
If the halo mass extends well beyond the virial radius, $\ct \gg 1$, 
the potential approaches
\be
U(r) \simeq \frac{c\mu}{(2-\alpha)}  
\left[ \left( \frac{x}{1+x} \right) ^{2-\alpha} -1 \right]\, \Vv^2 .
\ee 
Note that the potential as quoted in \equ{potential} is for 
$\Rt \rightarrow \infty$.

\smallskip
The velocity dispersion can be obtained under the assumption that the
phase-space distribution function depends only on energy, 
namely that the velocity dispersion tensor is isotropic.
The radial velocity dispersion has to satisfy the Jeans equation (or
hydrostatic equation), namely
\be
\frac{\dd (\rho \sigma_r^2)}{\dd r} = -\rho \frac{\dd U}{\dd r} .
\ee
If one assumes that $\Rt \rightarrow \infty$ and the boundary condition is
$\rho \sigma_r^2 = 0$ as $r \rightarrow \infty$,
one can obtain the velocity dispersion by performing the integral
\be
\sigma_r^2(r) = \frac{1}{\rho(r)} 
\int_r^\infty  \frac{M(y) \rho(y)}{y^2} {\dd} y .
\ee
This integral is expressed analytically 
in \equ{dispersion} following \citet{zhao96}.  
Already for $n=k=1$ it involves a sum of five terms, which we therefore
avoid spelling out here.
Explicit expressions for this case are given in 
equations 7-11 of \citet{tremaine94}.

\subsection{Two-parameter Potential for $\tgamma=3$ and $\beta=2$}
\label{sec:beta2}

We will see in \se{sims} that among the family of analytic profiles the case
$k=1$ with $n=2$ provides the most natural match to the shape of simulated DM 
profiles, with only two free parameters, $\alpha$ and $c$.
The density profiles are 
\be
\rho(r) = \frac{\rn}{x^\alpha\, (1+x^{1/2})^{2(3.5-\alpha)}} , 
\ee
\be
\brho(r) = \frac{\rb}{x^\alpha\, (1+x^{1/2})^{2(3-\alpha)}} .
\ee
The mass, velocity and force profiles are given in \equs{brho} to (\ref{eq:F})
with $\tgamma=3$ and $\beta=2$.

\smallskip
The potential, based on \equ{potential} but assuming that the density
profile is truncated at $\Rt=(\ct/c)\Rv$, is
\be
\begin{aligned}
U(r) =& -\Vt^2   -2 c \mu \Vv^2 \times \\
      &\left(
     \frac{\chi_{\ct}^{2(2-\alpha)} -\chi^{2(2-\alpha)}}{2(2-\alpha)}
    -\frac{\chi_{\ct}^{2(2-\alpha)+1} -\chi^{2(2-\alpha)+1}}{2(2-\alpha)+1}
  \right) ,
\end{aligned}
\ee
where
\be
\chi(x) = \frac{x^{1/2}}{1+x^{1/2}} , \quad \chi_{\ct}=\chi(x=\ct) . 
\ee
The term involving $\chi(\ct)$ ensures that $U(\Rt)=-\Vt^2$,
and the normalization factor in front of the big brackets guarantees that
$\dd U/\dd r = -F(r)$.


\smallskip
An analytic expression for the velocity dispersion profile
can be obtained from \equ{dispersion}.
It is an elaborate sum of many terms, which we avoid spelling out here.

\subsection{Three-Parameter Double Profiles}
\label{sec:double}

Any linear combination of the analytic profiles will naturally 
also have an analytic potential, and with more free parameters the fit 
to simulations can be made as good as desired.\footnote{\citet{schaller15} 
successfully used 
a linear combination of two functions for $\rho$, while here we use such a 
combination for $\brho$, keeping the analytic nature of the profiles.}
We will see that this may not
be necessary for a fit of the $\beta=2$ model in the range $(0.01-1)\Rv$,
but it may be useful for the $\beta=1$ model if an excellent
fit is desired at all radii in this range. 
The simplest option is a sum of two $\tgamma=3$ profiles with
the same $\alpha$ and $\beta$ but different concentrations, $c_1$ and $c_2$, 
namely three free parameters.
The number of free parameters is the same as in the single profile of
\equ{brho} with $\beta$ fixed and $\tgamma$ free, and we will see that the 
quality of the fit is also similar, except that for the double profile 
we also have an analytic potential.

\smallskip
We adopt the linear combination
\be
\brho(r) = \frac{\rba}{x_1^{\alpha} (1+x_1^{1/n})^{n(3-\alpha)}}
         + \frac{\rbb}{x_2^{\alpha} (1+x_2^{1/n})^{n(3-\alpha)}} , 
\label{eq:double}
\ee
where $x_i= c_i r/\Rv$ for $i=1,2$.
With the choice $c_1 > c_2$, the first and second terms are made to dominate
the inner and outer halo, respectively.
The value of $\alpha$ for the second component may be less important, so we 
let it be the same as $\alpha$ of the first component, thus keeping the number 
of free parameters at three.
The normalization coefficients $\rbi$ are determined such that the
fractional contribution of each component to $\brho$ at $\Rv$ is
$f_i$ ($f_1+f_2=1$), namely
\be
\rbi = f_i \brhov c_i^\alpha (1+c_i)^{3-\alpha} .
\ee
The values of $f_i$ are to be decided in advance, before the functional form
is used to match different simulated or observed profiles, so they should not
be regarded as additional free parameters. In order to chose fiducial values
for $f_i$, we perform in \se{sims} experimental fits to simulated profiles
where we do allow $f_i$ to vary.  We find that best fits to $\brho(r)$ are
obtained with $f_1$ in the range $0.1-0.5$, and therefore adopt hereafter
$f_1=0.33$ and $f_2=0.67$ as our fiducial fixed values.
We also note that the choice $f_1=f_2=0.5$ works slightly better when fitting
$\rho(r)$ rather than $\brho(r)$.
As long as $f_1$ is comparable to or slightly smaller than $f_2$,
their exact values do not make a significant difference and should be regarded
as fine tuning.

\smallskip
The associated profiles of local density, mass, velocity squared, force, 
potential and velocity-dispersion squared
are all analogous sums of two components.
The slope of $\brho(r)=\brho_1(r)+\brho_2(r)$ becomes
\be
\bs(r) = \frac{\brho_1(r)\, \bs_1(r) + \brho_2(r) \bs_2(r)}{\brho(r)} ,
\ee
where $\bs_i(r)$ are given in \equ{bslope} for the respective $x_i$ with
$\tgamma=3$ and $\beta=1$ or $2$.

\section{Fit to Simulations}
\label{sec:sims}

\subsection{The Simulations}
\label{sec:the_sims}

\subsubsection{General}

We use here three pairs of haloes from the NIHAO suite of zoom-in cosmological 
simulations \citep{wang15} at $z=0$.
The simulations are described, e.g., in \citet{tollet16,dutton16b}.
Each pair consists of simulations with and without baryons (``HYDRO" and ``DMO"
respectively), otherwise starting from the same initial conditions.
The six haloes thus span a range of profiles with a variety of 
inner cusps and cores.
The resolution allows an accurate recovery of the density profile at
$(0.01-0.02)\Rv$, where the evolution between cusp and core is most pronounced.

\smallskip
The standard flat $\Lambda$CDM cosmology was assumed, with the
Planck parameters \citep{planck14} 
($H_0= 67.1 \hmpc$, $\omm=0.3175$, 
$\oml=1-\omm=0.6825$, $\omb=0.0490$, $\sigma_8 = 0.8344$, $n=0.9624$).

\smallskip
The simulations were performed using the SPH code {\sc gasoline} 
\citep{wadsley04}, as modified by \citet{keller14} to reduce the formation 
of blobs and improve mixing.
The treatment of cooling via hydrogen, helium and metal-lines in a
uniform ultraviolet ionizing background is described in \citet{shen10}.
The star formation recipe is described in \citet{stinson06}.
The thermal stellar feedback, which is the main driver of evolution in the 
inner-halo profile, includes an early phase of winds and 
photoionization from massive stars, and a later epoch starting $4\Myr$ after
the star formation, when the first supernovae explode and dominate the feedback
thereafter.
The \citet{chabrier03} IMF is used.
Stars in the mass range $(8-40)\Msun$ eject an energy of $10^{51}\erg$ 
and metals into the interstellar medium surrounding stars.
Supernova feedback is implemented
using the blast-wave formalism described in \citet{stinson06}.  
To avoid rapid radiative cooling in the dense gas 
receiving the energy, cooling of gas particles inside the blast region is 
delayed for $\sim 30 \Myr$.  

\smallskip
The DM haloes were identified using the MPI+OpenMP hybrid halo finder
\texttt{AHF}\footnote{http://popia.ft.uam.es/AMIGA}
\citep{knollmann09,gill04}. \texttt{AHF} locates local over-densities
in an adaptively smoothed density field as prospective halo
centers. The virial masses of the haloes are defined as the masses
within a sphere containing $\Delta_{\rm v}=200$ times the cosmic critical
matter density, $\rho_{\rm crit}=3H(z)^2 / 8 \pi G$.

\smallskip
The galaxies produced in the NIHAO simulations match the main observational
constraints, including the Tully-Fisher relation, the stellar to
halo mass ratio, the main sequence of star-forming galaxies
\citep{wang15,wang17},
outflows through the CGM and metallicity \citep{gutcke17},
the presence of bulge-less disks \citep{obreja16},
the velocity function of ``too big to fail" dwarf galaxies \citep{dutton16a},
and the presence of a wide range of inner-halo
profiles ranging from cusp to core \citep{tollet16,dutton16b}.
This is encouraging in terms of the potential validity of the star-formation 
and feedback subgrid recipes and the resultant DM density profiles in the inner
halo. Our only concern here is that the six simulated profiles are  
representative of the variety of real halo profiles.

\subsubsection{Measuring the Profiles}
\label{sec:prof}

The mass profile is obtained by sorting the DM particles by their distance from
the halo center, yielding a rather smooth mass profile.
The mass profile is binned into points equally spaced in $\log r$, with spacing
of $1/35$ dex, namely about 70 points in the range of interest $(0.01-1)\Rv$.
The profiles of $\brho(r)$ and $V(r)$ are computed straightforwardly at these
grid points.
The local density profile $\rho(r)$ is obtained by a smooth derivative of the 
mass profile using a Savitzky-Golay filter \citep{savitzky64}, 
with a second-degree polynomial and a window size of $\sim 10$ bins. 
The smoothing is applied beyond the radius range of interest to avoid edge
effects.
The logarithmic slope profiles of $\brho$ and $\rho$ are obtained by similar
smooth derivatives using the same filter.

\smallskip
We consider the safe, reliable and interesting range of the profiles 
to be $(0.01-1)\Rv$, but
also show extended profiles from the simulations
below $0.01\Rv$ and out to $2\Rv$.
The actual gravitational softening radius of the simulation 
is typically a factor $2-3$ smaller than $0.01\Rv$, and the choice of
$0.01\Rv$ as a safe convergence radius for the NIHAO simulations is
justified in \citet[][section 2.4]{tollet16}.

\subsection{Fitting the Simulations}
\label{sec:fit_sims}

\subsubsection{Method}
\label{sec:fit_method}

We use the 3x2 simulated haloes at $z=0$
to evaluate the ability of the new analytic 
profiles to match the variety of realistic halo profiles, especially in the
inner halo, and to rank the relative goodness of fit among these profiles.
The halo masses in the hydro simulations are $2.7\times 10^{10}$, 
$1.3\times 10^{11}$ and $9.4\times 10^{11}\msun$.\footnote{These simulated 
galaxies 
are termed in \citep{wang15} g2.63e10, g2.19e11 and g8.06e11, respectively, 
referring to the halo masses from the low-resolution box from which the haloes 
where chosen.} 
We refer here to the simulations according to their log halo masses,
namely 10D, 11D and 12D for the DMO simulations
and 10H, 11H and 12H for the hydro simulations.

\smallskip
In all cases, the DMO profiles are cuspy, $\bsi \simeq 1.1-1.4$,
and can be well fit by the NFW profile and especially by the Einasto profile.
Their response to gas inflow and outflow is described in \citet{tollet16}.
In the high-mass halo 12H, where feedback-driven outflows are negligible
(producing a relatively high stellar-to-halo mass ratio 
$\Ms/\Mv =4.75\times 10^{-2}$),
the baryons lead to a contraction of the inner halo and thus to a steepening 
of the inner cusp in the halo density profile of the hydro simulation, 
from $\bsi =1.1$ to $1.3$.
In the intermediate-mass halo 11H, where there are intense episodes of inflow,
partly recycled, and the feedback-driven outflows are very effective 
(yielding a lower $\Ms/\Mv=7.08\times 10^{-3}$),
the baryons lead to a significant expansion of the inner halo, 
flattening the cusp to a core in the hydro simulation,
from $\bsi=1.3$ to $0.2$.
In the low-mass halo, where gas ejection is efficient and it suppresses 
the inflow such that the SFR becomes lower 
(with a very low $\Ms/\Mv=1.78 \times 10^{-3}$), 
the baryons lead to a weaker expansion, and a partial flattening of the cusp,
from $\bsi=1.4$ to $0.6$.

\begin{table*}
\centering

\setlength\tabcolsep{2.8pt}

\begin{tabular}{c>{\raggedright}p{0.3cm}c>{\raggedright}p{0.3cm}cc>{\centering}p{0.3cm}cc>{\centering}p{0.3cm}cc>{\centering}p{0.3cm}cc>{\centering}p{0.3cm}cc>{\centering}p{0.3cm}cc>{\centering}p{0.3cm}cc}
\hline 
\noalign{\vskip0.05cm}
 &  & {\footnotesize{}halo \#} &  &  &  &  & \multicolumn{2}{c}{{\footnotesize{}1}} &  & \multicolumn{2}{c}{{\footnotesize{}2}} &  & \multicolumn{2}{c}{{\footnotesize{}3}} &  & \multicolumn{2}{c}{{\footnotesize{}4}} &  & \multicolumn{2}{c}{{\footnotesize{}5}} &  & \multicolumn{2}{c}{{\footnotesize{}6}}\tabularnewline[0.05cm]
\noalign{\vskip0.05cm}
 &  & {\footnotesize{}name} &  &  &  &  & \multicolumn{2}{c}{{\footnotesize{}10D}} &  & \multicolumn{2}{c}{{\footnotesize{}11D}} &  & \multicolumn{2}{c}{{\footnotesize{}12H}} &  & \multicolumn{2}{c}{{\footnotesize{}12D}} &  & \multicolumn{2}{c}{{\footnotesize{}10H}} &  & \multicolumn{2}{c}{{\footnotesize{}11H}}\tabularnewline[0.05cm]
\noalign{\vskip0.05cm}
 &  & {\footnotesize{}$\bar{s}_{1}$} &  &  &  &  & \multicolumn{2}{c}{{\footnotesize{}1.4}} &  & \multicolumn{2}{c}{{\footnotesize{}1.3}} &  & \multicolumn{2}{c}{{\footnotesize{}1.3}} &  & \multicolumn{2}{c}{{\footnotesize{}1.1}} &  & \multicolumn{2}{c}{{\footnotesize{}0.6}} &  & \multicolumn{2}{c}{{\footnotesize{}0.2}}\tabularnewline[0.05cm]
\hline 
\hline 
\noalign{\vskip0.05cm}
{\footnotesize{}\#} &  & {\footnotesize{}model} &  & \multicolumn{2}{c}{{\footnotesize{}params}} & \multicolumn{18}{c}{}\tabularnewline[0.05cm]
\cline{1-6} 
\noalign{\vskip0.1cm}
\multirow{4}{*}{{\footnotesize{}a1}} & \multirow{4}{0.3cm}{} & \multirow{4}{*}{{\footnotesize{}$\bar{\gamma}$ free, $\beta=1$}} & \multirow{4}{0.3cm}{} & {\scriptsize{}$\Delta$} & {\scriptsize{}$\Delta\bar{s}_{1}$} &  & \textbf{\scriptsize{}0.008} & \textbf{\scriptsize{}0.08} &  & \textbf{\scriptsize{}0.010} & \textbf{\scriptsize{}0.12} &  & \textbf{\scriptsize{}0.012} & \textbf{\scriptsize{}0.12} &  & \textbf{\scriptsize{}0.018} & \textbf{\scriptsize{}0.16} &  & \textbf{\scriptsize{}0.008} & \textbf{\scriptsize{}0.08} &  & \textbf{\scriptsize{}\uline{0.008}} & \textbf{\scriptsize{}\uline{-0.06}}\tabularnewline
 &  &  &  & {\scriptsize{}$c$} & {\scriptsize{}$c_{{\rm m}}$} &  & {\scriptsize{}7.6} & {\scriptsize{}8.3} &  & {\scriptsize{}7.6} & {\scriptsize{}5.5} &  & {\scriptsize{}23.4} & {\scriptsize{}8.3} &  & {\scriptsize{}6.2} & {\scriptsize{}4.2} &  & {\scriptsize{}26.6} & {\scriptsize{}7.1} &  & {\scriptsize{}21.6} & {\scriptsize{}4.6}\tabularnewline
 &  &  &  & {\scriptsize{}$\alpha$} & {\scriptsize{}$\bar{s}_{1}$} &  & {\scriptsize{}1.3} & {\scriptsize{}1.5} &  & {\scriptsize{}1.3} & {\scriptsize{}1.4} &  & {\scriptsize{}1.1} & {\scriptsize{}1.4} &  & {\scriptsize{}1.1} & {\scriptsize{}1.3} &  & {\scriptsize{}-0.1} & {\scriptsize{}0.7} &  & {\scriptsize{}-0.6} & {\scriptsize{}0.1}\tabularnewline
 &  &  &  & {\scriptsize{}$\bar{\gamma}$} & {\scriptsize{}$\bar{s}_{2}$} &  & {\scriptsize{}2.7} & {\scriptsize{}2.6} &  & {\scriptsize{}2.5} & {\scriptsize{}2.4} &  & {\scriptsize{}2.3} & {\scriptsize{}2.3} &  & {\scriptsize{}2.6} & {\scriptsize{}2.4} &  & {\scriptsize{}2.5} & {\scriptsize{}2.4} &  & {\scriptsize{}2.6} & {\scriptsize{}2.4}\tabularnewline[0.05cm]
\hline 
\noalign{\vskip0.05cm}
\multirow{4}{*}{{\footnotesize{}a2}} & \multirow{4}{0.3cm}{} & \multirow{4}{*}{{\footnotesize{}$\bar{\gamma}$ free, $\beta=2$}} & \multirow{4}{0.3cm}{} & {\scriptsize{}$\Delta$} & {\scriptsize{}$\Delta\bar{s}_{1}$} &  & \textbf{\scriptsize{}\uline{0.005}} & \textbf{\scriptsize{}\uline{0.03}} &  & \textbf{\scriptsize{}0.007} & \textbf{\scriptsize{}0.09} &  & \textbf{\scriptsize{}0.010} & \textbf{\scriptsize{}0.08} &  & \textbf{\scriptsize{}\uline{0.014}} & \textbf{\scriptsize{}0.12} &  & \textbf{\scriptsize{}\uline{0.003}} & \textbf{\scriptsize{}\uline{0.01}} &  & \textbf{\scriptsize{}0.012} & \textbf{\scriptsize{}-0.14}\tabularnewline
 &  &  &  & {\scriptsize{}$c$} & {\scriptsize{}$c_{{\rm m}}$} &  & {\scriptsize{}6.2} & {\scriptsize{}8.3} &  & {\scriptsize{}6.4} & {\scriptsize{}5.2} &  & {\scriptsize{}328} & {\scriptsize{}8.3} &  & {\scriptsize{}3.4} & {\scriptsize{}3.9} &  & {\scriptsize{}442} & {\scriptsize{}6.6} &  & {\scriptsize{}157} & {\scriptsize{}4.3}\tabularnewline
 &  &  &  & {\scriptsize{}$\alpha$} & {\scriptsize{}$\bar{s}_{1}$} &  & {\scriptsize{}0.9} & {\scriptsize{}1.4} &  & {\scriptsize{}0.8} & {\scriptsize{}1.4} &  & {\scriptsize{}-1.1} & {\scriptsize{}1.4} &  & {\scriptsize{}0.7} & {\scriptsize{}1.2} &  & {\scriptsize{}-5.2} & {\scriptsize{}0.6} &  & {\scriptsize{}-4.5} & {\scriptsize{}0.1}\tabularnewline
 &  &  &  & {\scriptsize{}$\bar{\gamma}$} & {\scriptsize{}$\bar{s}_{2}$} &  & {\scriptsize{}3.3} & {\scriptsize{}2.6} &  & {\scriptsize{}3.0} & {\scriptsize{}2.4} &  & {\scriptsize{}2.5} & {\scriptsize{}2.3} &  & {\scriptsize{}3.3} & {\scriptsize{}2.4} &  & {\scriptsize{}2.9} & {\scriptsize{}2.5} &  & {\scriptsize{}3.1} & {\scriptsize{}2.5}\tabularnewline[0.05cm]
\hline 
\noalign{\vskip0.05cm}
\multirow{4}{*}{{\footnotesize{}b1}} & \multirow{4}{0.3cm}{} & \multirow{4}{*}{{\footnotesize{}$\bar{\gamma}=3,$ $\beta=1$}} & \multirow{4}{0.3cm}{} & {\scriptsize{}$\Delta$} & {\scriptsize{}$\Delta\bar{s}_{1}$} &  & \textbf{\scriptsize{}0.015} & \textbf{\scriptsize{}0.16} &  & \textbf{\scriptsize{}0.020} & \textbf{\scriptsize{}0.24} &  & \textbf{\scriptsize{}0.031} & \textbf{\scriptsize{}0.37} &  & \textbf{\scriptsize{}0.022} & \textbf{\scriptsize{}0.26} &  & \textbf{\scriptsize{}0.041} & \textbf{\scriptsize{}0.42} &  & \textbf{\scriptsize{}0.042} & \textbf{\scriptsize{}0.25}\tabularnewline
 &  &  &  & {\scriptsize{}$c$} & {\scriptsize{}$c_{{\rm m}}$} &  & {\scriptsize{}3.6} & {\scriptsize{}7.3} &  & {\scriptsize{}2.2} & {\scriptsize{}4.2} &  & {\scriptsize{}1.8} & {\scriptsize{}4.9} &  & {\scriptsize{}2.5} & {\scriptsize{}3.7} &  & {\scriptsize{}6.0} & {\scriptsize{}5.4} &  & {\scriptsize{}7.2} & {\scriptsize{}4.1}\tabularnewline
 &  &  &  & {\scriptsize{}$\alpha$} & {\scriptsize{}$\bar{s}_{1}$} &  & {\scriptsize{}1.5} & {\scriptsize{}1.6} &  & {\scriptsize{}1.5} & {\scriptsize{}1.5} &  & {\scriptsize{}1.6} & {\scriptsize{}1.7} &  & {\scriptsize{}1.3} & {\scriptsize{}1.4} &  & {\scriptsize{}0.8} & {\scriptsize{}1.0} &  & {\scriptsize{}0.2} & {\scriptsize{}0.5}\tabularnewline
 &  &  &  &  & {\scriptsize{}$\bar{s}_{2}$} &  &  & {\scriptsize{}2.7} &  &  & {\scriptsize{}2.5} &  &  & {\scriptsize{}2.5} &  &  & {\scriptsize{}2.5} &  &  & {\scriptsize{}2.7} &  &  & {\scriptsize{}2.7}\tabularnewline[0.05cm]
\hline 
\noalign{\vskip0.05cm}
\multirow{4}{*}{{\footnotesize{}b2}} & \multirow{4}{0.3cm}{} & \multirow{4}{*}{{\footnotesize{}$\bar{\gamma}=3,$ $\beta=2$}} & \multirow{4}{0.3cm}{} & {\scriptsize{}$\Delta$} & {\scriptsize{}$\Delta\bar{s}_{1}$} &  & \textbf{\scriptsize{}0.009} & \textbf{\scriptsize{}-0.03} &  & \textbf{\scriptsize{}0.007} & \textbf{\scriptsize{}0.08} &  & \textbf{\scriptsize{}0.018} & \textbf{\scriptsize{}0.24} &  & \textbf{\scriptsize{}0.015} & \textbf{\scriptsize{}\uline{0.07}} &  & \textbf{\scriptsize{}0.008} & \textbf{\scriptsize{}0.09} &  & \textbf{\scriptsize{}0.012} & \textbf{\scriptsize{}-0.17}\tabularnewline
 &  &  &  & {\scriptsize{}$c$} & {\scriptsize{}$c_{{\rm m}}$} &  & {\scriptsize{}22.9} & {\scriptsize{}9.5} &  & {\scriptsize{}6.7} & {\scriptsize{}5.2} &  & {\scriptsize{}4.4} & {\scriptsize{}6.4} &  & {\scriptsize{}9.2} & {\scriptsize{}4.2} &  & {\scriptsize{}123} & {\scriptsize{}6.2} &  & {\scriptsize{}238} & {\scriptsize{}4.3}\tabularnewline
 &  &  &  & {\scriptsize{}$\alpha$} & {\scriptsize{}$\bar{s}_{1}$} &  & {\scriptsize{}0.4} & {\scriptsize{}1.4} &  & {\scriptsize{}0.8} & {\scriptsize{}1.4} &  & {\scriptsize{}1.1} & {\scriptsize{}1.5} &  & {\scriptsize{}0.5} & {\scriptsize{}1.2} &  & {\scriptsize{}-2.5} & {\scriptsize{}0.7} &  & {\scriptsize{}-5.6} & {\scriptsize{}0.0}\tabularnewline
 &  &  &  &  & {\scriptsize{}$\bar{s}_{2}$} &  &  & {\scriptsize{}2.6} &  &  & {\scriptsize{}2.4} &  &  & {\scriptsize{}2.4} &  &  & {\scriptsize{}2.4} &  &  & {\scriptsize{}2.5} &  &  & {\scriptsize{}2.5}\tabularnewline[0.05cm]
\hline 
\noalign{\vskip0.05cm}
\multirow{4}{*}{{\footnotesize{}c1}} & \multirow{4}{0.3cm}{} & \multirow{4}{*}{{\footnotesize{}x2\enskip{} $\bar{\gamma}=3$, $\beta=1$}} & \multirow{4}{0.3cm}{} & {\scriptsize{}$\Delta$} & {\scriptsize{}$\Delta\bar{s}_{1}$} &  & \textbf{\scriptsize{}0.005} & \textbf{\scriptsize{}0.04} &  & \textbf{\scriptsize{}0.008} & \textbf{\scriptsize{}0.11} &  & \textbf{\scriptsize{}0.012} & \textbf{\scriptsize{}0.16} &  & \textbf{\scriptsize{}0.015} & \textbf{\scriptsize{}0.13} &  & \textbf{\scriptsize{}0.006} & \textbf{\scriptsize{}0.02} &  & \textbf{\scriptsize{}0.011} & \textbf{\scriptsize{}-0.13}\tabularnewline
 &  &  &  & {\scriptsize{}$c_{1}$} & {\scriptsize{}$c_{{\rm m}}$} &  & {\scriptsize{}11.2} & {\scriptsize{}8.9} &  & {\scriptsize{}7.3} & {\scriptsize{}5.2} &  & {\scriptsize{}9.6} & {\scriptsize{}8.3} &  & {\scriptsize{}7.3} & {\scriptsize{}4.2} &  & {\scriptsize{}26.7} & {\scriptsize{}6.6} &  & {\scriptsize{}23.2} & {\scriptsize{}3.8}\tabularnewline
 &  &  &  & {\scriptsize{}$c_{2}$} & {\scriptsize{}$\bar{s}_{1}$} &  & {\scriptsize{}2.5} & {\scriptsize{}1.4} &  & {\scriptsize{}1.3} & {\scriptsize{}1.4} &  & {\scriptsize{}0.9} & {\scriptsize{}1.4} &  & {\scriptsize{}1.6} & {\scriptsize{}1.2} &  & {\scriptsize{}5.0} & {\scriptsize{}0.6} &  & {\scriptsize{}5.7} & {\scriptsize{}0.1}\tabularnewline
 &  &  &  & {\scriptsize{}$\alpha$} & {\scriptsize{}$\bar{s}_{2}$} &  & {\scriptsize{}1.2} & {\scriptsize{}2.6} &  & {\scriptsize{}1.2} & {\scriptsize{}2.4} &  & {\scriptsize{}1.2} & {\scriptsize{}2.3} &  & {\scriptsize{}1.1} & {\scriptsize{}2.4} &  & {\scriptsize{}-0.3} & {\scriptsize{}2.6} &  & {\scriptsize{}-0.9} & {\scriptsize{}2.6}\tabularnewline[0.05cm]
\hline 
\noalign{\vskip0.05cm}
\multirow{4}{*}{{\footnotesize{}c2}} & \multirow{4}{0.3cm}{} & \multirow{4}{*}{{\footnotesize{}x2\enskip{} $\bar{\gamma}=3$, $\beta=2$}} & \multirow{4}{0.3cm}{} & {\scriptsize{}$\Delta$} & {\scriptsize{}$\Delta\bar{s}_{1}$} &  & \textbf{\scriptsize{}0.009} & \textbf{\scriptsize{}-0.03} &  & \textbf{\scriptsize{}\uline{0.007}} & \textbf{\scriptsize{}\uline{0.07}} &  & \textbf{\scriptsize{}\uline{0.010}} & \textbf{\scriptsize{}\uline{0.04}} &  & \textbf{\scriptsize{}0.015} & \textbf{\scriptsize{}\uline{0.07}} &  & \textbf{\scriptsize{}0.006} & \textbf{\scriptsize{}0.03} &  & \textbf{\scriptsize{}0.012} & \textbf{\scriptsize{}-0.17}\tabularnewline
 &  &  &  & {\scriptsize{}$c_{1}$} & {\scriptsize{}$c_{{\rm m}}$} &  & {\scriptsize{}22.9} & {\scriptsize{}9.5} &  & {\scriptsize{}16.9} & {\scriptsize{}5.2} &  & {\scriptsize{}313} & {\scriptsize{}7.3} &  & {\scriptsize{}9.2} & {\scriptsize{}4.2} &  & {\scriptsize{}5.2e4} & {\scriptsize{}6.2} &  & {\scriptsize{}237} & {\scriptsize{}4.3}\tabularnewline
 &  &  &  & {\scriptsize{}$c_{2}$} & {\scriptsize{}$\bar{s}_{1}$} &  & {\scriptsize{}22.9} & {\scriptsize{}1.4} &  & {\scriptsize{}5.9} & {\scriptsize{}1.3} &  & {\scriptsize{}29.7} & {\scriptsize{}1.3} &  & {\scriptsize{}9.2} & {\scriptsize{}1.2} &  & {\scriptsize{}1.5e4} & {\scriptsize{}0.6} &  & {\scriptsize{}238} & {\scriptsize{}0.0}\tabularnewline
 &  &  &  & {\scriptsize{}$\alpha$} & {\scriptsize{}$\bar{s}_{2}$} &  & {\scriptsize{}0.4} & {\scriptsize{}2.6} &  & {\scriptsize{}0.6} & {\scriptsize{}2.4} &  & {\scriptsize{}-2.2} & {\scriptsize{}2.4} &  & {\scriptsize{}0.5} & {\scriptsize{}2.4} &  & {\scriptsize{}-62.3} & {\scriptsize{}2.5} &  & {\scriptsize{}-5.6} & {\scriptsize{}2.5}\tabularnewline[0.05cm]
\hline 
\end{tabular}

\caption{A summary of the fits of six models to the six simulated haloes.
The haloes are ordered by the slope $\bsi$ at $0.015\Rv$.
The models, with $\beta=1,2$, are (a) the three-parameter flexible model 
with free $\tgamma$,
(b) the analytic model with $\tgamma=3$ and two free parameters,
and (c) the double model with $\tgamma=3$ and three free parameters.
The quality of the fit is estimated by $\Delta$ and $\Delta \bar{s}_1$ (in bold
face),
the rms log residuals in $(0.01-1)\Rv$ and the deviation of $\bar{s}_1$ from
the simulated value.
The other entries are the free parameters of the functional form 
($c$, $\alpha$, $\tgamma$)
and the associated physical parameters 
($c_{\rm m}$, related to the velocity maximum, and $\bar{s}_1$ and $\bar{s}_2$,
the slopes at $0.015\Rv$ and at $\Rv$.
The best fit in every column is marked by an underline.
}
\label{tab:w1}
\end{table*}

\smallskip
We note in the example simulations shown below that in the DMO simulations
the profile of the {\it slope} of $\brho(r)$ in $(0.01-1)\Rv$ is well fit by 
a power law, indicating that the Einasto profile will be a good fit.
However, the slope profiles in the hydro simulations tend to
deviate from a power law, so the Einasto profile is not expected to be a good
fit.
 
\smallskip
We fit each of the new analytic functional forms discussed in the previous two 
sections to each of the simulated profiles. 
The fit is performed on the binned profile of $\brho$ with no further
smoothing.
The fitting method is Levenberg-Marquardt least squares
\citep{levenberg44,marquardt63}.
The rms of the residuals of $\log\brho$, denoted $\Delta$, is used to evaluate
a {\it relative} global goodness of fit.
The absolute value of $\Delta$ is sensitive to how smooth the target simulated
profile is (namely the resolution of the simulations and the binning procedure
for the profile), so it should mostly serve for comparing the performance of
different models in fitting target profiles that were measured in the same way
rather than for an absolute goodness of fit.

\smallskip
The binning is important in order to allow us to focus the fit on a desired
specific radius range, such as the cusp-core region,
while still obtaining a sensible fit in other regions.
With the bins equally spaced in log radius, the effective weight assigned to 
the inner halo is larger than it would have been in a maximum-likelihood
fit performed with equal weights to each particle.
When we wish to assign an enhanced weight to a given region, 
we may assign enhanced weights, by a factor $w$, to the data points in 
bins that lie in this region. 
The value of $w$ is decided in advance according to the focus of the study, 
e.g., $w=1$ when an overall accurate fit is desired, or $w>1$ in the inner 
halo when the focus is on the cusp-core region. This weighting procedure 
makes only a minor difference.  We should stress again that the binning and
the nonuniform weighting allow only a relative goodness-of-fit estimate, 
not an absolute one.

\subsubsection{Summary of Fit Results}

\begin{figure*}
\vskip 10.2cm
\includegraphics{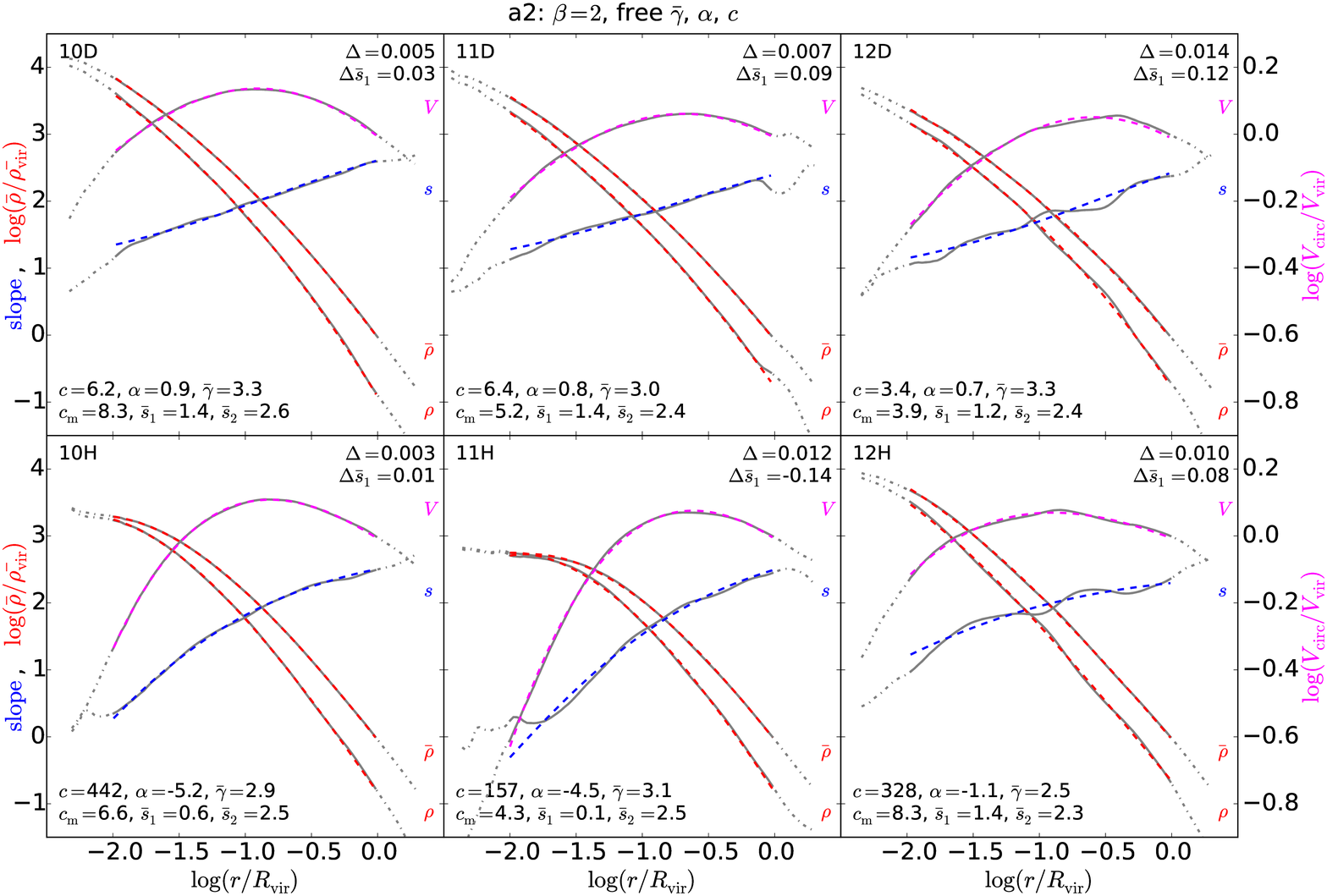}
\caption{
Best-fit model (dashed) versus the simulated profiles (solid).
Shown are the profiles of $\brho$, $\rho$, $V$ and the $\brho$ slope $\bar{s}$.
This figure is for the flexible model a2,
with $\tgamma$ free and $\beta=2$ (three free parameters).
The fits are excellent, with $\Delta =0.003-0.014$ dex and 
$\Delta \bar{s}_1=0.01-0.14$.
\Fig{a1} shows the same for model a1, with $\beta=1$, where the fits are
slightly less good, except for halo 11H with the flat core.
}
\label{fig:a2}
\end{figure*}
\begin{figure*}
\vskip 10.2cm
\includegraphics{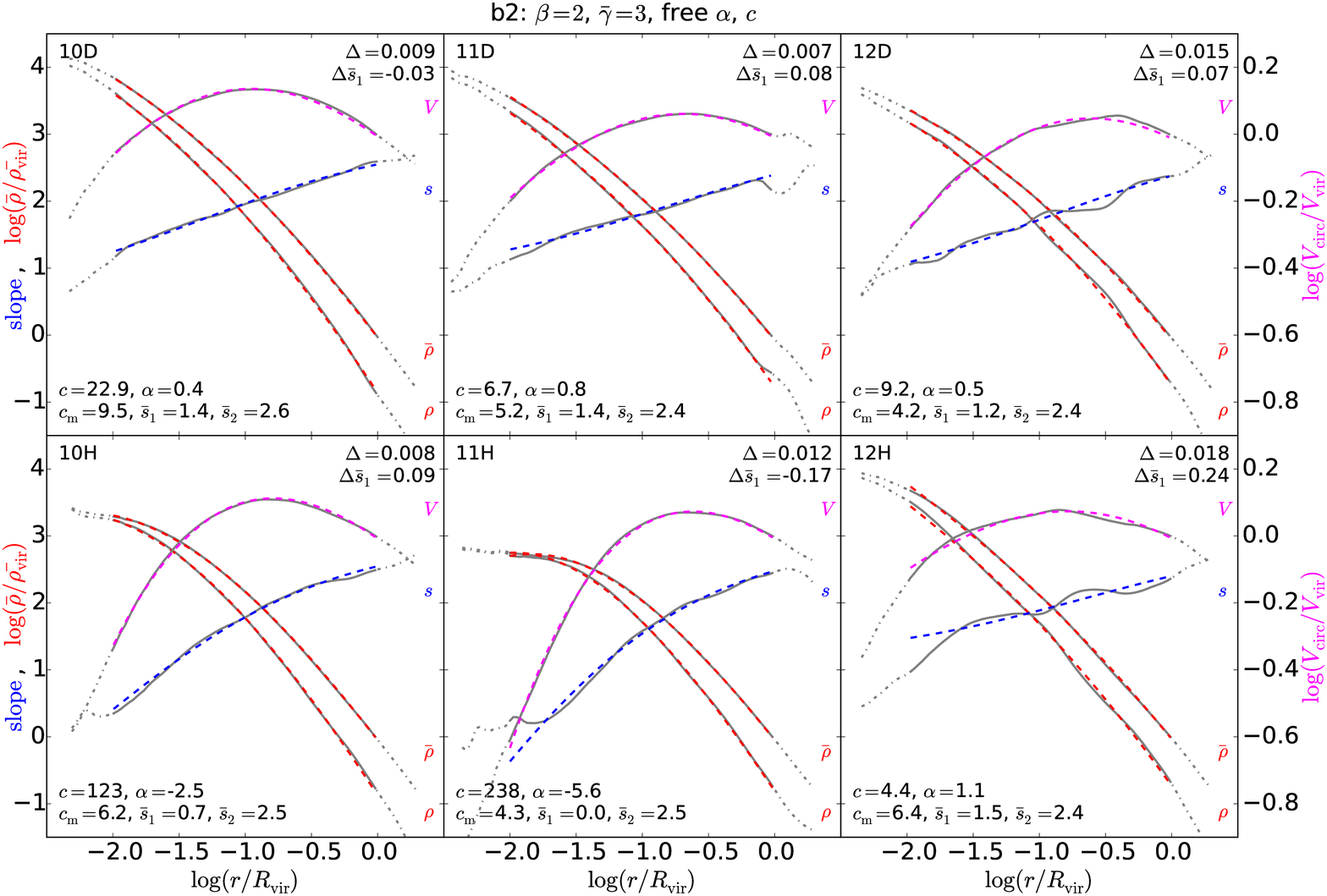}
\caption{
Same as \fig{a2} but for the fully analytic model b2, where
$\tgamma=3$ and $\beta=2$ (two free parameters).
The fit is still excellent, with 
$\Delta=0.007-0.018$ dex and $\Delta \bar{s}_1=0.03-0.24$.
\Fig{b1} shows the same for model b1, with $\beta=1$, where the fits are
less good for all haloes.
}
\label{fig:b2}
\end{figure*}
\begin{figure*}
\vskip 10.2cm
\includegraphics{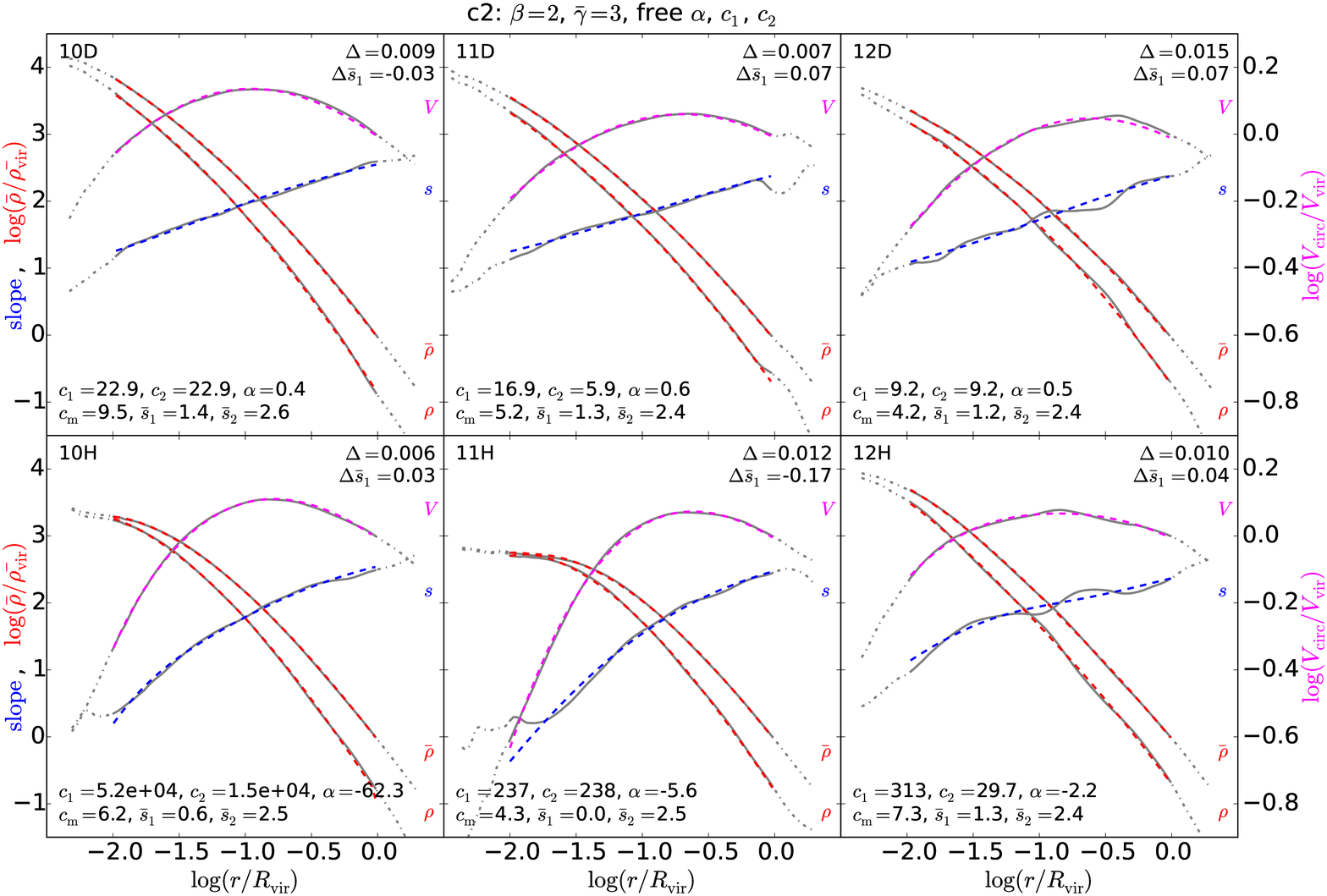}
\includegraphics{figs/lsfit_brho_dbl_b2_g3.eps}
\caption{
Same as \fig{a2} but for the fully analytic double model c2, where
$\tgamma=3$ and $\beta=2$ (three free parameters).
The fit is excellent, with
$\Delta=0.006-0.015$ dex and $\Delta \bar{s}_1=0.03-0.17$.
\Fig{c1} shows the same for model c1, with $\beta=1$, where the fits are
comparable.
}
\label{fig:c2}
\end{figure*}

The results of the fits of the 6 models to the 6 target simulated profiles 
with uniform weights ($w=1$) are summarized in \tab{w1}. 
The fits of the three models with $\beta=2$ are shown in 
\figs{a2} to \ref{fig:c2},
and the analogous models with $\beta=1$ are shown in \figs{a1} to \ref{fig:c1}.
In \se{w10} we bring the equivalent table and figures for fits with a high
weight of $w=10$ at
$r=(0.01-0.03)\Rv$, the most interesting region of cusp-to-core transition.

\smallskip
The table lists the haloes in columns and the models in rows.
The haloes are marked 1 to 6, ordered by the inner slope 
$\bsi$ at $0.015\Rv$ from steep to flat, which is quoted. 
The crude halo masses and D or H help identifying the haloes, 
from the DMO and hydro simulations respectively.
The models, each in two versions with $\beta=1$ and $2$, are
(a) the three-parameter free-$\tgamma$ model with maximum flexibility and
mass-velocity analytic profiles,
(b) the two-parameter $\tgamma=3$ model with analytic potential, and
(c) the three-parameter double model with $\tgamma=3$ and 
an analytic potential. 

\smallskip
The entries for each model-halo pair are first of all two estimates of the
goodness of fit (in bold face): 
$\Delta$ is the overall rms of log residuals of $\brho(r)$ in the range 
$(0.01-1)\Rv$, and 
$\Delta \bar{s}_1$ is the deviation of the inner slope $\bar{s}_1$ in the 
model from the simulated halo profile.
Then, quoted in the left column are the values of the best-fit parameters
of the functional form (e.g., $c$, $\alpha$ and $\tgamma$), 
and in the right column the values of the parameters with physical meaning
($c_{\rm m}$, $\bar{s}_1$, and $\bar{s}_2$ - the slope at $\Rv$).
The best fitting model in terms of $\Delta$ or $\Delta \bar{s}_1$
for each halo (namely in each column) is marked by an underline.


\smallskip 
The fits are in general quite good, with the rms log residual 
$\Delta \sim 0.01$, ranging from below one percent to a few percent.
The model inner slope matches the true value with deviations 
$\bar{s}_1 \sim 0.1$, ranging from $0.01$ to $0.4$.
For $\beta=2$ typically $\Delta \lsim 0.01$ and $\bar{s}_1 \lsim 0.1$.

\smallskip 
For the cusped haloes, no. 1 to 4, the best-fit models are with $\beta=2$.
The three-parameter models a2 and c2 are naturally slightly better in most
cases, but the two-parameter model b2 is comparable in two haloes and not far 
behind in the other two.  
The best fit for the semi-cored halo \#5 is by model a2, and the best fit to the
cored halo \#6 is actually by model a1, with $\beta=1$. 

\smallskip 
Overall, when an analytic potential is required, namely $\tgamma=3$, 
best accuracy is provided by model c2, namely with $\beta=2$ and
three free parameters. 
However, for simplicity, model b2 provides fits that are comparable and
almost as accurate, with only two free parameters (and $\beta=2$).
For the most accurate fits without a requirement for an analytic potential,
the models with free $\tgamma$ provide slightly better fits, and better
with $\beta=2$ than $\beta=1$, except for the cored halo. 
When an extremely accurate fit is desired,
and only the velocity-mass profile is required to be analytic, one can appeal
to the similar model with a free $\beta$.

\smallskip 
When an enhanced weight is assigned to the core-cusp region, $w=10$,
we see in \tab{w10} that
the fit of the inner slope is naturally better, with $\Delta \bar{s}_1$
values typically ranging from $0.00$ to $0.05$
compared to $0.01-0.24$ with $w=1$. 
This is at the expense of the overall fit, which is typically of comparable 
or lower quality with respect to the $w=1$ case, sometimes by a factor of 
$\sim 2$ in $\Delta$.

\smallskip 
While the values of the free parameters in the functional form vary
significantly from model to model for the same halo, the physical parameters
robustly characterize each halo independently of the model used.
For example, $c_{\rm m}$ typically varies by less than $\pm 10\%$ from model to 
model for a given halo.  
The values of $\bar{s}_1$ and $\bar{s}_2$ typically vary by less than 
$\pm 0.1$ for a given halo.  
Among the haloes, the values of $c_{\rm m}$ vary from $3.8$ to $9.5$, 
and they are weakly correlated with the inner slope $\bar{s}_1$.
The outer slope of $\brho(r)$ at $\Rv$ is robust at $\bar{s}_2 = 2.5 \pm 0.1$.

\smallskip
\Figs{a2} to \ref{fig:c2} show the best fits of the three models with $\beta=2$
to the simulated profiles, focusing on the range $(0.01-1)\Rv$.
The analogous models with $\beta=1$ are shown in \figs{a1} to \ref{fig:c1}.
We show here results with uniform weights, $w=1$, 
while in \se{w10} we show examples with $w=10$ at $r=(0.01-0.03)\Rv$, the most
interesting region of cusp-to-core transition.
Shown are the profiles of mean density $\brho(r)$, local density $\rho(r)$, 
circular velocity $V(r)$, and mean-density slope $\bar{s}(r)$.
The best-fit values of the free parameters, and the two measures of quality of
fit, same as in \tab{w1}, are quoted in the figures. 
We discuss these figures here.

\subsubsection{Free-$\tgamma$ Profiles with $\beta=1,2$, Models a1 and a2}
\label{sec:fit_sims_a}

\Figs{a2} and \ref{fig:a1} refer to the 
three-parameter flexible model of \se{flex_brho}, \equ{brho},  
with the outer slope $\tgamma$ free, in addition to $\alpha$ and $c$.
Recall that this model has analytic expressions for the density and
mass-velocity profiles but not for the potential profile.
For illustrative purposes
we fix $\beta$ at either $1$ or $2$, though any value of $\beta$ can be used
here.

\smallskip
The fits are excellent at all radii in all cases for the two values of $\beta$.
The rms deviations within $(0.01-1)\Rv$ are 
$\Delta=0.003-0.014$ dex for $\beta=2$ and 
$\Delta=0.008-0.018$ dex for $\beta=1$.
The inner slope deviations are 
$\Delta \bar{s}_1=0.01-0.14$ and $0.08-0.16$ respectively.
We learn that $\beta=2$, in general, provides better fits than $\beta=1$.

\smallskip
As expected, the useful parameters for characterizing the cusp-core are 
$\bsi$ (not $\alpha$) and to a certain extent $\cm$ 
(though it involves $\alpha$ and $\tgamma$).
The values of $\bsi$ are in the 
range $1.4-0.1$, while $\alpha$ ranges from $0.9$ to large negative values
that have no physical interpretation.
The values of $\cm$ are limited to the relatively narrow range 
$3.9-8.3$, while the values of $c$ can become extremely large, and therefore 
lack a physical meaning.
The values of $\bs_{\Rv}$ for model a2 are stable in the narrow range 
$2.3-2.6$, while the values of $\tgamma$ are somewhat larger and they span
a broader range, $2.5-3.3$. 
 
\smallskip
We conclude that the three-parameter model with free $\tgamma$ and $\beta=2$
can be very useful in matching the profiles in all cases, where analytic 
density and mass-velocity profiles are desired but an analytic potential is
not required. This function may be useful in particular
for the study of the outer profile, near and outside $\Rv$,
which could be affected by tides as a function of the halo
environment and is expected to vary with the accretion rate onto the halo
\citep{diemer14}.

\subsubsection{Analytic Profiles with $\tgamma=3$, Models b1 and b2}
\label{sec:fit_sims_b}

\Fig{b2} and \fig{b1} refer to the simple, two-parameter, $\tgamma=3$,
fully analytic models b2 and b1,
with $\beta=2$ and $\beta=1$ respectively.
The free parameters are $\alpha$ and $c$.
Model b1 has a somewhat simpler analytic expression for the potential,
and a much simpler expression for the velocity dispersion,
but model b2 is a better fit to the simulated haloes, and we therefore focus on
it here.
 
\smallskip
The profile with $\beta=2$ turns out to more naturally match the shape of the
simulated profiles in the middle halo, both for the cases of cusps and cores.
In particular, it allows to capture the non-power-law slope profile in the
cases of a core, and the slopes near $\Rv$.
The fit for $\beta=2$ has $\Delta \sim 0.007-0.018$ and
$\Delta \bar{s}_1 = 0.03-0.24$. This is excellent, though naturally not as 
good as the three-parameter model a2.  

\smallskip
As expected in \se{physical}, with $\beta=2$ 
the values of the parameters $\alpha$ and $c$
are not meaningful for the radius range of interest (this is true for models
a2 and c2 as well).
While in the cuspy cases
these values are in the same ball park as in the other models ($\alpha=0.4-1.1$
and $c=4.4-22.9$, in the cases of a flatter core $\alpha$ is negative and large
($-2.5$ and $-5.6$) and $c$ is very large ($123-238$), meaning that $\rs$
falls outside the range of interest, below $0.01\Rv$.
This makes the model profile at very small radii well below $0.01\Rv$ 
irrelevant to what real haloes are likely to look like on such small scales.
In fact, in this model the density profile at very small radii is rising
with radius and the density vanishes as $r \rightarrow 0$.
The main virtue of this model is the excellent match to the variety of halo
profiles at $(0.01-1)\Rv$ with a fully analytic profile and only 
two free parameters. 
However, we stress again that for the properties of physical
interest one should appeal to quantities such as $\bsi$, $\bsv$ 
and $\cm$, and one should not extrapolate this profile to radii
well outside the range where the fit was performed.

\smallskip
We conclude that the $\tgamma=3$ with $\beta=2$ is an excellent fully analytic
profile for fitting the range $(0.01-1)\Rv$ in haloes of a variety of cusps and
cores, given the quality of the fits and having only two free parameters.
Its disadvantages for some purposes are the possible large deviations of the
profile shape from realistic haloes well outside the range of interest,
and the somewhat less simple analytic expressions for the potential and
especially for the velocity dispersion compared to the model b1 with
$\beta=1$.

\smallskip
On the other hand, the fit with $\beta=1$ in \fig{b1}, especially
in the cases of a core, tends to overestimate
the density in the middle halo, near the velocity peak,
and to overestimate the inner slope, with
$\Delta \sim 0.015-0.042$ and
$\Delta \bar{s}_1 = 0.16-0.42$.
As can be seen in \se{w10}, the match in the inner slope can be improved by
enhanced weighting ($w=10$) in the inner halo, but this comes at the expense of
increasing the deviation in the middle halo. It seems that $\beta=1$ does not
really capture the shapes of the profiles in the middle halo.

\smallskip
With $\beta=1$,
the value of the inner asymptotic slope $\alpha$ is similar to the slope of
interest $\bsi$ for the cuspy profiles, but $\alpha$ underestimates
$\bsi$ by $0.2-0.3$ for the flatter inner profiles. This implies that for
low values of $\alpha$, in order to evaluate the core profile one should
appeal to $\bsi$ rather than to $\alpha$ even when $\beta=1$.
The slope of the local $\rho(r)$ at $0.015\Rv$ is larger than $\bsi$ of
$\brho(r)$ by the $\Delta s$ given in \equ{Ds}.

\smallskip
With $\beta=1$ the values of $c$ are not ridiculously large 
(as they are for $\beta=2$).
The low values of $c$ compared to model a1 or to the NFW case
are due to the need to compensate for the
enforced $\tgamma=3$ at $r \gg \Rv$ (corresponding to $\gamma=4$ for
$\rho(r)$), so the meaning of $c$ is not as straightforward as in these
Note that $c$ gets larger in the cored cases, namely $c$ tends to be
anti-correlated with $\alpha$.
On the other hand, $\cm$ is somewhat correlated with $\bsi$, 
so it may serve as an additional characteristic of the cusp-core.

\smallskip
We conclude that the single $\tgamma=3$ and $\beta=1$ model can be used to
study the cusp-core when a simple analytic potential is needed,
but only if a very accurate fit in the middle halo is not required.
We will see in \se{fit_sims_double}
that the fit becomes excellent overall when a double such model is used.
Its advantage is that the analytic expression for the potential is very simple,
and the analytic expression for the velocity dispersion is manageable.

\subsubsection{Double Profiles with $\tgamma=3$, Models c1 and c2}
\label{sec:fit_sims_double}

\Fig{c2} and \fig{c1} refer to   
the fully analytic double profiles of \equ{double},
with $\tgamma=3$ and with $\beta$ either $2$ or $1$ in models c2 and c1
respectively.
The three free parameters are $\alpha$, $c_1$ and $c_2$.
We use fixed weights for the two components, $f_1=0.33$ and $f_2=0.67$ 
(giving best fit when fitting $\brho$), and uniform weighting ($w=1$).

\smallskip
With three free parameters,
the fits are excellent at all radii in all cases,
with  
$\Delta=0.005-0.015$ dex and
$\Delta \bar{s}_1 = 0.03-0.17$ for the two values of $\beta$.
The double profiles capture the inner halo, the peak velocity, 
and the outskirts.
For $\beta=2$, the quality of fit of the double profile is comparable to 
that of the single profile, but for $\beta=1$ the double profile represents a
significant improvement over the single profile. 

\smallskip
Again, the useful parameter for characterizing the cusp-core is $\bsi$, 
not $\alpha$, which is materialized at a smaller radius, not relevant to the
core region of interest.
The values of $\alpha$ can be negative, especially in the cored haloes,
and with $\beta=2$ they could become very large.

\smallskip
The values of $c_1$ and $c_2$ are not straightforward to interpret, and for
$\beta =2$ and cored haloes they become extremely large.
The more physical concentration is in a rather narrow range,
$\cm=3.8-9.5$,
and is has some correlation with $\bsi$, so it
can also be used to characterize the cusp-core.

\smallskip
We conclude that the double $\tgamma=3$ models are both very well 
suited for studying the evolution in the cusp-core where an analytic potential 
is needed, with the $\beta=1$ double model being a significant improvement 
over the corresponding single model. 
For these double models,
the quality of fits for $\beta=2$ and $\beta=1$ are similar. 
This argues in favor of preferring the double model with $\beta=1$ (c1),
because the analytic expressions for the potential and velocity dispersion 
are simpler.
However, recall that the single model with $\beta=2$ (b2) is as accurate, 
and it involves only two free parameters.

\begin{figure*}
\vskip 13.0cm
\includegraphics{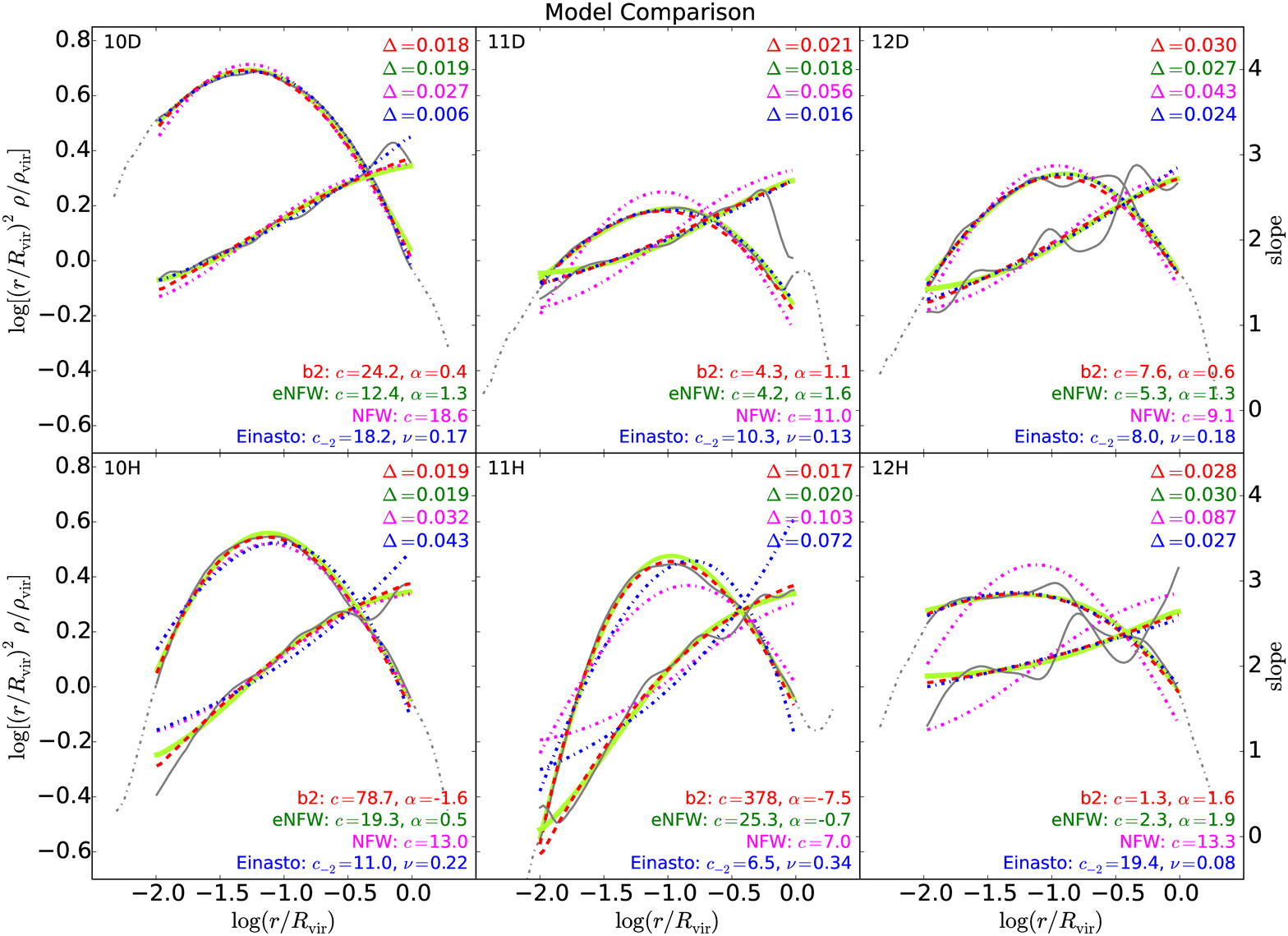}
\caption{
A comparison of models versus the simulated halo profiles, 
where the best-fit is applied to $\rho(r)$,
showing $r^2\rho(r)$ and the slope of $\rho(r)$.
Compared are the new analytic profile b2 ($\beta=2$, $\tgamma=3$; red), 
the eNFW profile with $\alpha$ free ($\beta=1$, $\gamma=3$; green), 
and the Einasto profile (blue),
each with two free parameters.
Also shown is the one-parameter NFW fit ($\alpha=1$).
The best fits of the analytic b2 model and the eNFW profile are similar,
and they are both excellent in quality with low values of $\Delta$ for all 
haloes. 
The NFW profile is fine for some of the cuspy haloes, but it becomes a poor
approximation for other cuspy haloes and for the cored haloes.
The Einasto profile does as well as eNFW and the analytic profile
in the cuspy cases, but it fails in the cases with a flatter cusp-core, 
where the density slope profile deviates from a power law.
}
\label{fig:compare_rho}
\end{figure*}

\subsubsection{Fits with Enhanced Weighting in the Inner Halo}

In \se{w10} we present in \tab{w10} and \figs{a2_w10} to \ref{fig:c2_w10}
the analogous fits of models to simulated haloes
with enhanced weighting of $w=10$ in the inner halo, $(0.01-0.03)\Rv$.
The fits naturally improve in the cusp-core region for all models and all
haloes, with  
$\Delta \bar{s}_1 = 0.00-0.03, 0.01-0.10, 0.00-0.05$
for models a,b,c respectively.
The values of $\Delta \bar{s}_1 = 0.07-0.10$ are limited to model b1,
but even these represent very small deviations.

\smallskip
The overall fit is somewhat less good than with equal weighting ($w=1$), 
with 
$\Delta = 0.004-0.024, 0.009-0.083, 0.007-0.024$
for models a,b,c respectively.
Again, the values of $\Delta = 0.054-0.083$ are limited to model b1.
In general, the global rms deviations of order 0.01 dex are sensible.

\smallskip
We conclude that when the focus is on the fit in the cusp-core region, one can
benefit from applying the fit with enhanced weights at $(0.01-0.03)\Rv$.

\subsection{Comparison with Popular Non-analytic Models}
\label{sec:comparison}

It would be worthwhile to compare the fits 
between the new models and other popular models which do not necessarily have
analytic expressions for mass-velocity and potential.
We compare our model b2 
with an extended NFW profile (eNFW)
and with the Einasto profile, all having two free parameters.

\smallskip
By eNFW we refer to \equ{flex} with $\beta=1$ and $\gamma=3$, where $\alpha$ is
free in addition to $c$. We also show for comparison the one-parameter
NFW profile where $\alpha=1$.

\smallskip
The Einasto density profile 
\citep{einasto65,navarro04}
is
\be
\rho(r) = \rho_{-2} \exp \left( -\frac{2}{\nu} (x^\nu -1)\right) ,
\quad x=\frac{r}{r_{-2}} ,
\label{eq:einasto}
\ee
where $r_{-2}$ is where the slope is $-2$ and the density is $\rho_{-2}$.
This is the 3D analog of the Sersic profile used to match the stellar surface
density profiles of galaxies.
This Einasto density profile has a power-law slope profile,
\be
s(r) = -\frac{\dd \log \rho}{\dd \log r}=2\left( \frac{r}{r_{-2}} \right)^\nu ,
\ee
and the best fit to DMO simulated haloes yields $\nu \simeq 0.17$
\citep{navarro04,gao08,duffy08,dutton14}. 

\smallskip
\Fig{compare_rho} shows the different best-fit models versus the simulated
profiles. Here, we show the profiles of local density $r^2\rho(r)$ and the 
slope of $\rho(r)$ (rather than the analogous quantities for $\brho(r)$ shown
in all other figures), because for $\rho(r)$
one has analytic expressions for all models.
The fit is performed here on the simulated $\rho(r)$, 
derived from $M(r)$ via a procedure that involves certain smoothing as 
described in \se{prof}, with equal weights in log-spaced radii.

\smallskip
The best fits of the analytic b2 model and the eNFW profile are rather
similar, and they are both excellent in quality for all haloes,
with $\Delta=0.018-0.030$.
In the cuspy cases, the Einasto profile does as well and even slightly better
than the eNFW and analytic profiles, with $\Delta=0.006-0.027$,
but it fails in the cases with a flatter cusp-core,
where the density slope profile deviates from a power law, with
$\Delta=0.043-0.072$.
The NFW profile is fine for some of the cuspy haloes, though not as good as the
two-parameter profiles. NFW is a poor approximation for other cuspy haloes, 
and for the cored haloes, with $\Delta=0.032-0.103$.

\smallskip
We conclude that the matches of our new two-parameter profile to the simulated 
profiles are similar in quality to other existing profiles with a similar 
number of free parameters. 
The virtue
of 
 the new profile is the analytic expressions for the mass-velocity,
potential, and velocity dispersion. 
The analytic models with three parameters provide slightly better fits, 
comparable to other profiles with three free parameters, with the advantage of
analytic mass-velocity profiles.

\section{Conclusion}
\label{sec:conc}

Our proposed functional form for the {\it mean} density profile of spherical
 DM haloes, 
with a varying asymptotic inner slope $\alpha$, is based on \equ{brho}.
By expressing the mean density (rather than the local density) in simple 
analytic terms, the mass, velocity and force profiles are automatically 
expressed analytically, and the local density profile is easily derived.
The most flexible functional form involves four parameters, $\alpha$, $\beta$,
$\tgamma$ and $c$.
In principle, $\beta$ and $\tgamma$ can vary, but in practice $\beta \simeq 2$ 
yields excellent fits, and $\beta=1$ can also provide good fits in certain 
models.

\smallskip
When the asymptotic outer slope of $\brho(r)$ is fixed at $\tgamma=3$, 
and $\beta$ is a natural number such that the asymptotic slope of $\rho(r)$
is $\gamma=\tgamma+\beta^{-1}$,
there are also {\it analytic} expressions for the {\it potential} and 
velocity dispersion. 
These provide a new useful tool  %
for theoretical studies of halo evolution and for constructing
model haloes.
With the introduction of a free concentration parameter $c$, the $\tgamma=3$
profiles have the flexibility for matching the outer slopes of DM-halo 
profiles.

\smallskip
The six models tested here, all with either $\beta=2$ or $\beta=1$,
are 
(a) a flexible model with three free parameters $\alpha$, $c$ and $\tgamma$,
(b) a model with $\tgamma=3$ and two free parameters $\alpha$ and $c$,
and (c) a double profile, a sum of models as in b, with three free parameters
$\alpha$, $c_1$ and $c_2$. 
Models a have {\it analytic mass and velocity profiles}.
Models b and c have in addition {\it analytic expressions for the potential and
velocity-dispersion profiles}.

\smallskip
We evaluate the relative quality of these models by performing 
fits to profiles of six DM haloes from cosmological simulations without and 
with baryons, in which the inner profiles range form a steep cusp to a flat 
core. 

\smallskip
We find that
the best fits are provided by models a2, b2, and c1 or c2.
This says that in general $\beta=2$ captures better the shape of the profile
in the middle halo, but a double profile with $\beta=1$ can mimic a similar
shape in the middle halo.

\smallskip
If an excellent fit is desired at all radii, 
with analytic mass-velocity profiles but
without a need for an analytic potential,
and if the fit has to extend well beyond the virial radius,
the flexible model a2 with a free $\tgamma$ is a good choice.  
If an even better accuracy is desired, this model can be applied with a free
$\beta$.

\smallskip
If an {\it analytic potential} is required, the minimal model b2, with 
$\beta=2$ and only two free parameters, is our best choice.  
The encouraging finding is that model b2
provides fits almost as good as the three-parameter models. 
Therefore, there is no much gain in extending it to the double model c2.

\smallskip
Model b1 has somewhat simpler expressions for the potential and velocity
dispersion, but its fits to the simulated haloes are somewhat less accurate.
It is therefore the choice when the accuracy of the fits is not a major issue.
However,
if an excellent fit is desired, as well as simple analytic expressions,
the choice is the double model c1, with $\beta=1$.
 
\smallskip
We find that our analytic two-parameter model b2 matches the simulated profiles
as well as the popular eNFW and Einasto profiles, which have the same number of
parameters but no analytic expressions for mass-velocity and potential. 
In fact, model b2 does much better than Einasto in the case of cored haloes.
Our analytic double models do as well as other non-analytic
three-parameter models.

\smallskip
The free parameters in the functional form, the asymptotic slopes $\alpha$ and
$\tgamma$ and the concentration parameter(s), are not always useful for 
directly interpreting the shape of the profile in the range of interest,
$(0.01-1)\Rv$. This is true in particular for the models with $\beta=2$, 
where the profile well outside this radius range can be a very poor fit to 
the actual halo profiles.
A general warning is that an extrapolation of a best-fit model to outside
the fitting range is risky, and may be totally unrealistic, e.g., when
$\beta=2$.
 
\smallskip
The profile is characterized better by more physical parameters that can be 
derived from the free parameters of the functional form. The physical
parameters are, for example, the
actual slopes of $\brho(r)$ in the regions of
interest, e.g., $\bsi$ and $\bsv$, and the alternative concentration
parameter $\cm$, 
which refers to the radius where the slope of $\brho(r)$ is
$-2$ and where the velocity curve peaks.

\smallskip
We reiterate that the main purpose of this paper is to provide a new tool
for studying the evolution of dark-matter halo profiles, where there are 
fully {\it analytic} expressions for the mass and velocity profiles and
in particular for the gravitational {\it potential} and velocity dispersion
profiles.
For example, 
model b2, with $\tgamma=3$ and $\beta=2$, is being successfully used in an 
analytic study of the evolution of the 
inner halo profiles due to episodes of gas inflow and rapid outflow 
(Dekel et al, 2017, in prep.).

\section*{Acknowledgments}

We acknowledge stimulating discussions with Andi Burkert and Frank van den
Bosch.
This work was partly supported by the grants ISF 124/12, I-CORE Program of the
PBC/ISF 1829/12, BSF 2014-273, PICS 2015-18, GIF I-1341-303.7/2016,
and NSF AST-1405962.
The simulations were performed on the {\sc theo} cluster of the
Max-Planck-Institut f\"ur Astronomie and the {\sc hydra} cluster at the
Rechenzentrum in Garching; and the Milky Way supercomputer, funded by
the Deutsche Forschungsgemeinschaft (DFG) through Collaborative
Research Center (SFB 881) ``The Milky Way System'' (subproject Z2),
hosted and co-funded by the J\"ulich Supercomputing Center (JSC).



\appendix
\section{General Analytic Profiles}
\label{sec:zhao}

This appendix summarizes the results from \citet{zhao96} for the family of
local density profiles 
\be
\rho(x) \propto \frac{1}{x^{\alpha}\,(1+x^{1/n})^{n(3+k/n-\alpha)} } ,
\ee
where $n$ and $k$ are natural numbers.\footnote{We denote by $\alpha$ what
\citet{zhao96} denote as $\gamma$.}
Defining 
\be
\chi(x) = \frac{x^{1/n}}{1+x^{1/n}} ,
\ee
the density becomes
\be
\rho(x) \propto \chi^{-n\alpha} (1-\chi)^{3n+k} .
\ee

\smallskip
The mass profile is
\be
M(r) \propto \sum_{i=0}^{k-1} a_i \chi^{n(3-\alpha)+i} ,
\ee
where
\be
a_i =  \frac{n}{n(3-\alpha)+i}\, q(k-1,i) ,
\ee
and
\be
\begin{aligned}
q(i,j)
&=(-1)^j \frac{i!}{j!(1-j)!} , \quad i\geq j\geq 0 \\
&=0 , \quad {\rm otherwise} .
\end{aligned}
\ee
To be consistent with our notation,
one should normalize the mass and density profiles to obtain $M(\Rv)=\Mv$.

\smallskip
Assuming here that the halo density profile extends to infinity (while we
assume that it is truncated at $\Rt$),
and that the potential vanishes at infinity, 
the potential is
\be
U(r) \propto -\sum_{i=0}^{n+k-2} b_i\, S_{n(2-\alpha)+1}(\chi) ,
\label{eq:potential}
\ee
where
\be
b_i=n \sum_{j=0}^{i} q(n-1,j)\, a_{i-j} ,
\ee
and
\be
\begin{aligned}
S_i(\chi)
&=\frac{1-\chi^i}{i} , \quad i\neq 0 \\
&=-\log \chi , \quad i=0 .
\end{aligned}
\ee

\smallskip
The velocity dispersion is
\be
\sigma^2(r) \propto \frac{1}{\rho(r)}
\sum_{i=0}^{4n+2k-2} d_i\, S_{2n(1-\alpha)+i} (\chi) ,
\label{eq:dispersion}
\ee
where
\be
d_i=\sum_{j=0}^{i} e_i\, a_{i-j} ,
\ee
and
\be
e_i = n q(4n+k-1,i) .
\ee

\vfill\eject
\section{Fit to Simulations with $\beta=1$}
\label{sec:w1_more_figs}

\begin{figure*}
\vskip 10.2cm
\includegraphics{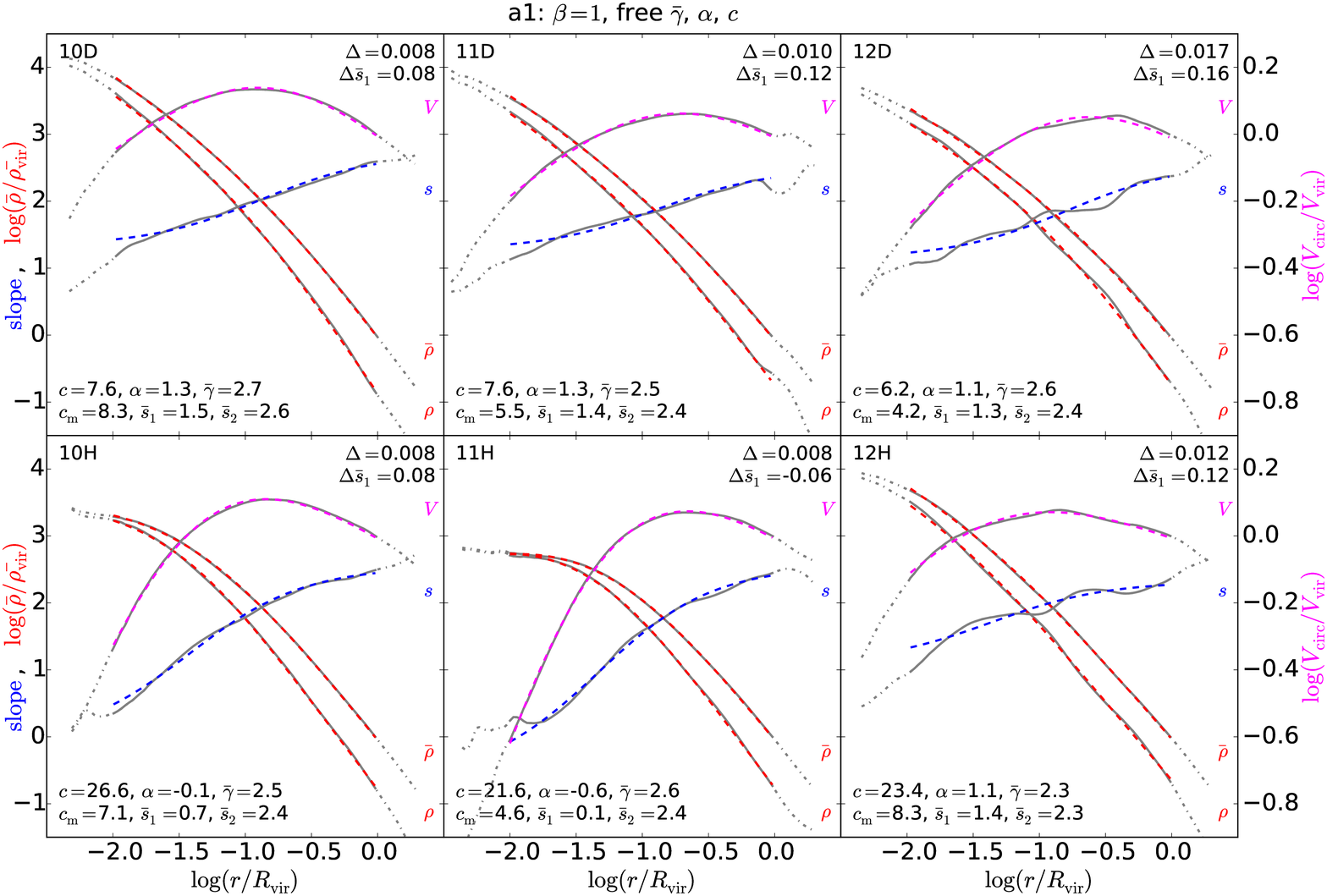}
\caption{
Same as \fig{a2}, for the flexible model, but for $\beta=1$ (model a1).
The fits are excellent, with
$\Delta=0.008-0.017$ and $\Delta \bar{s}_1=0.06-0.16$,
but not as good as the fits with $\beta=2$.
}
\label{fig:a1}
\end{figure*}

\begin{figure*}
\vskip 10.2cm
\includegraphics{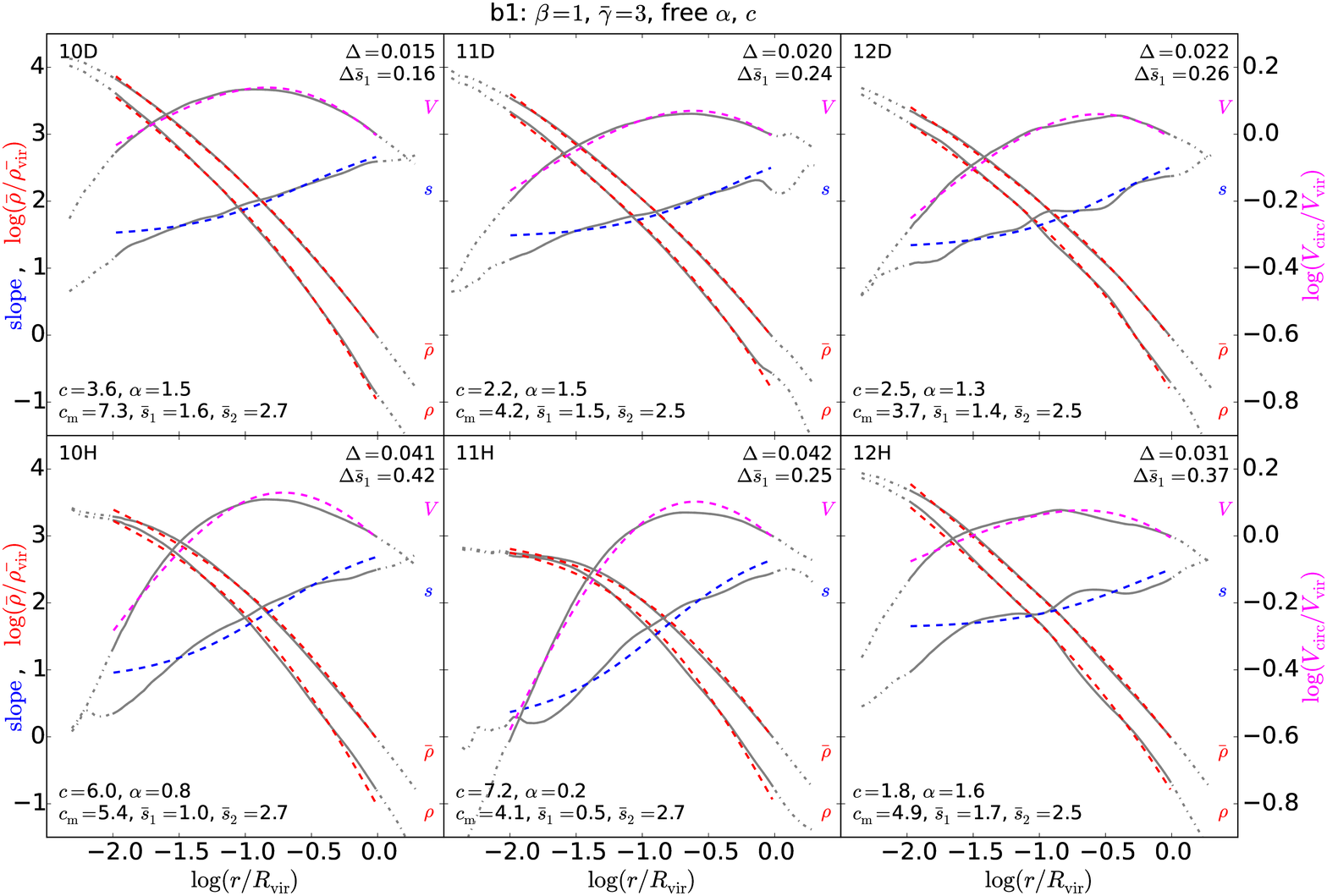}
\caption{
Same as \fig{b2}, for the two-parameter analytic model, but for $\beta=1$
(model b1).
The fits are fine in the cuspy haloes, with 
$\Delta=0.015-0.031$ and $\Delta \bar{s}_1=0.16-0.37$,
but less so in the middle halo in the cored cases, with
$\Delta=0.041-0.042$ and $\Delta \bar{s}_1=0.25-0.42$.
The $\beta=2$ model (b2) is significantly better.
}
\label{fig:b1}
\end{figure*}

\begin{figure*} 
\vskip 10.2cm
\includegraphics{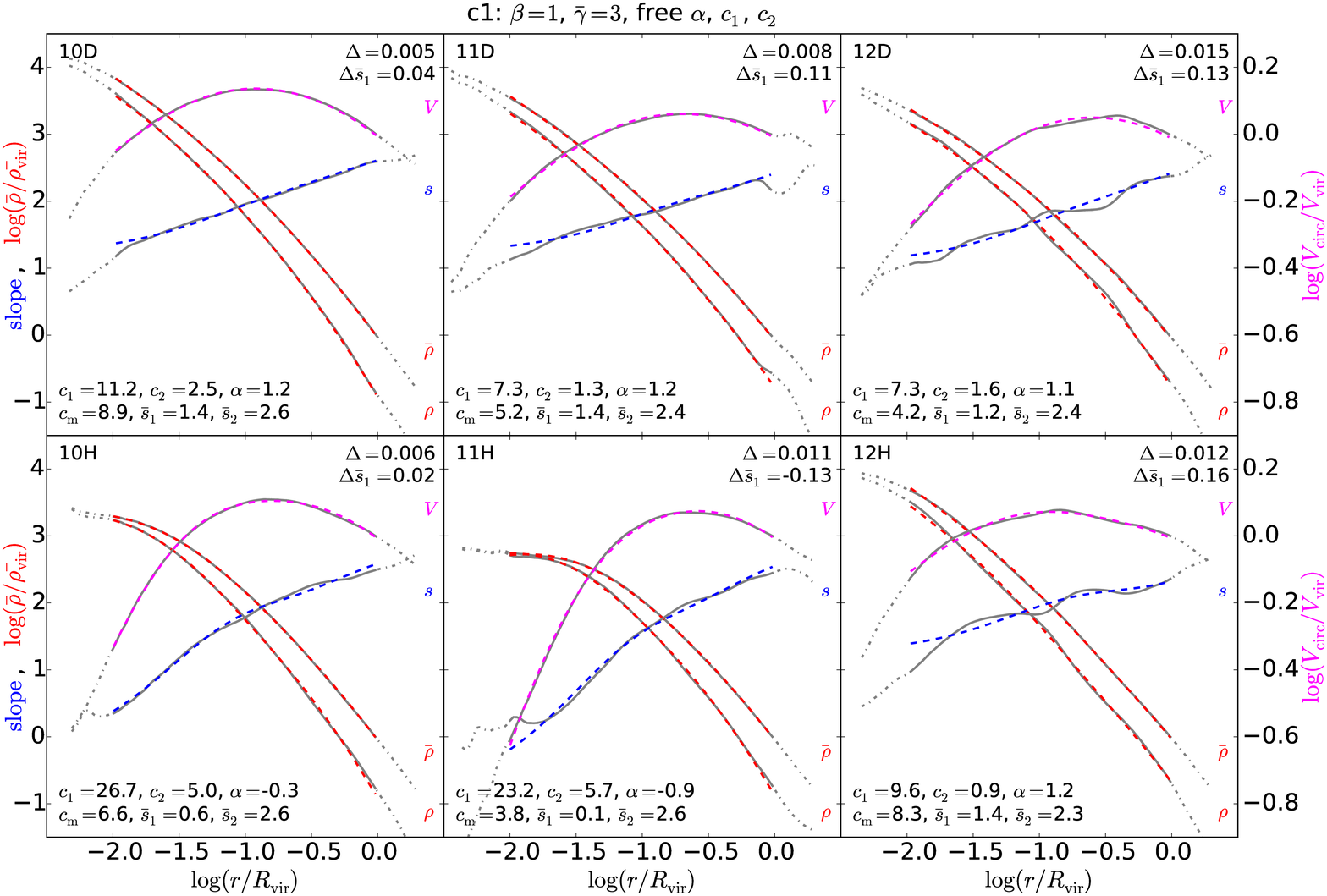} 
\caption{ 
Same as \fig{c2}, for the double analytic model, but for $\beta=1$ (model c1).
The fits are excellent, with
$\Delta=0.005-0.015$ and $\Delta \bar{s}_1=0.02-0.16$,
comparable to the fits of model c2 where $\beta=2$.
}
\label{fig:c1}
\end{figure*}

In \se{fit_sims} we present and discuss the fits of our new models to 
simulations. \tab{w1} summarizes the results of these fits, and \figs{a2} 
to \ref{fig:c2} help visualize the fits for the models with $\beta=2$ 
(models a2, b2 and c2). 
Here we complement this visual presentation in \figs{a1} to \ref{fig:c1}
which show the fits for the analogous models with $\beta=1$ 
(models a1, b1 and c1).
These fits are discussed in \se{fit_sims}.

\section{Fit to Simulations with Enhanced Weighting in the Inner Halo}
\label{sec:w10}

\begin{figure*}
\vskip 10.2cm
\includegraphics{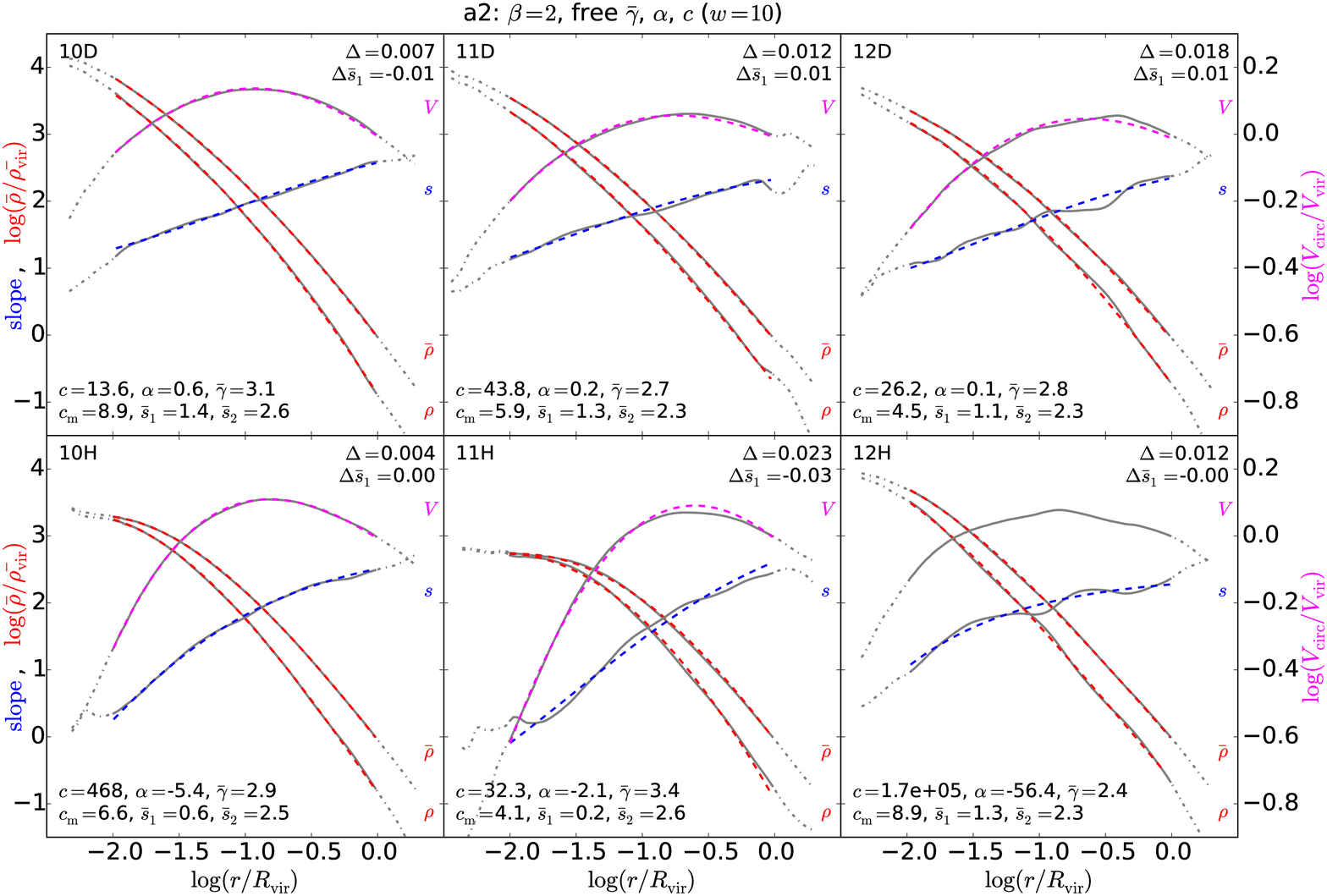}
\caption{
Same as \fig{a2}, for the flexible model with $\beta=2$ (model a2),
but with enhanced weighting $w=10$ in the fit at $(0.01-0.03)\Rv$.
The fits in the inner halo are excellent, with
$\Delta \bar{s}_1=0.00-0.03$. 
The overall fit is also good, $\Delta=0.004-0.023$, though less good than
with uniform weighting, especially in the cored halo 11H.
}
\label{fig:a2_w10}
\end{figure*}

\begin{figure*}
\vskip 10.2cm
\includegraphics{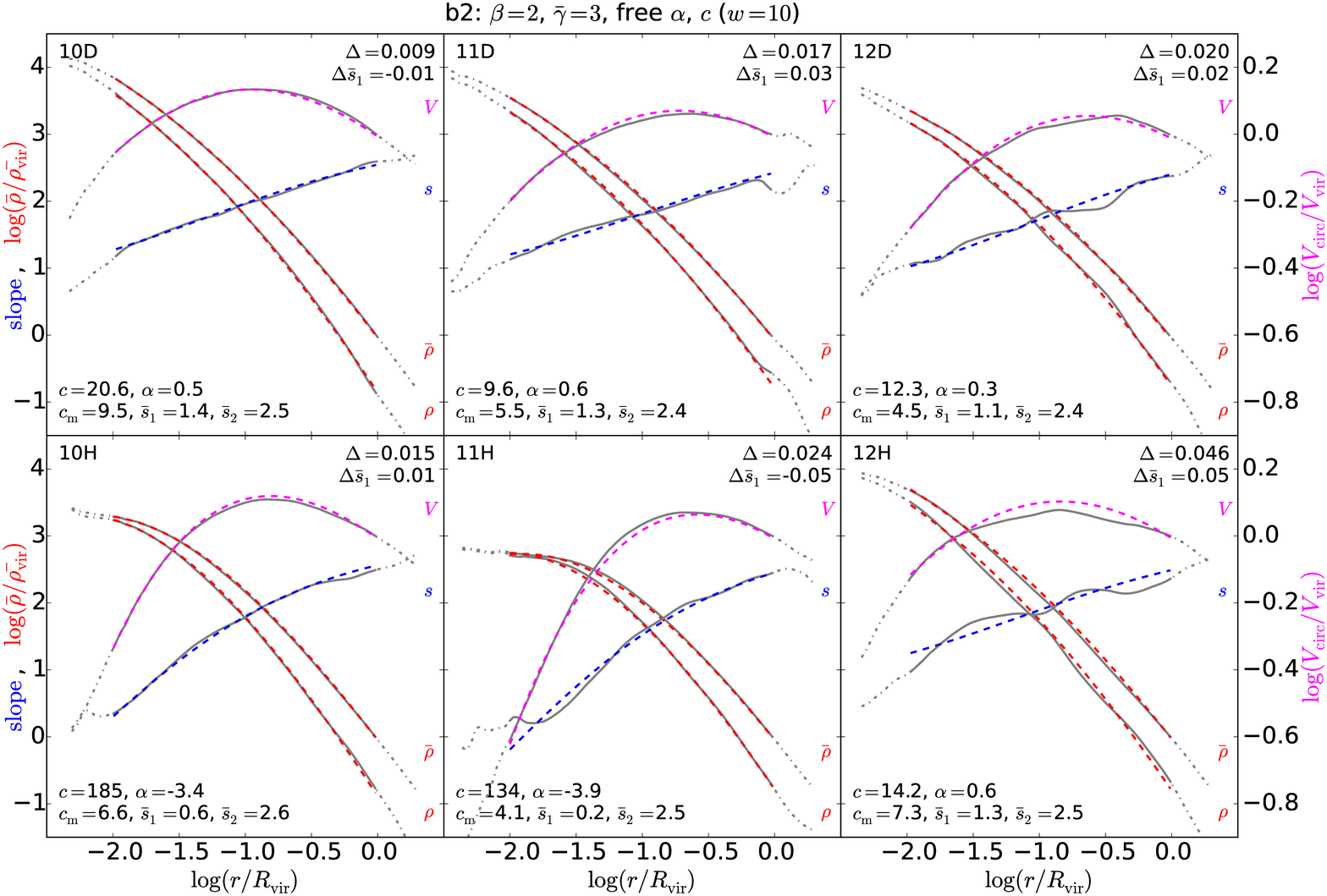}
\caption{
Same as \fig{b2}, for the two-parameter analytic model with $\beta=2$ (model
b2), 
but with enhanced weighting $w=10$ in the fit at $(0.01-0.03)\Rv$.
The fits in the inner halo are excellent, with
$\Delta \bar{s}_1=0.01-0.05$.
The overall fit is also fine, $\Delta=0.009-0.046$, 
though less good than with uniform weighting,
especially in the very cuspy halo 12H.
}
\label{fig:b2_w10}
\end{figure*}

\begin{figure*}
\vskip 10.2cm
\includegraphics{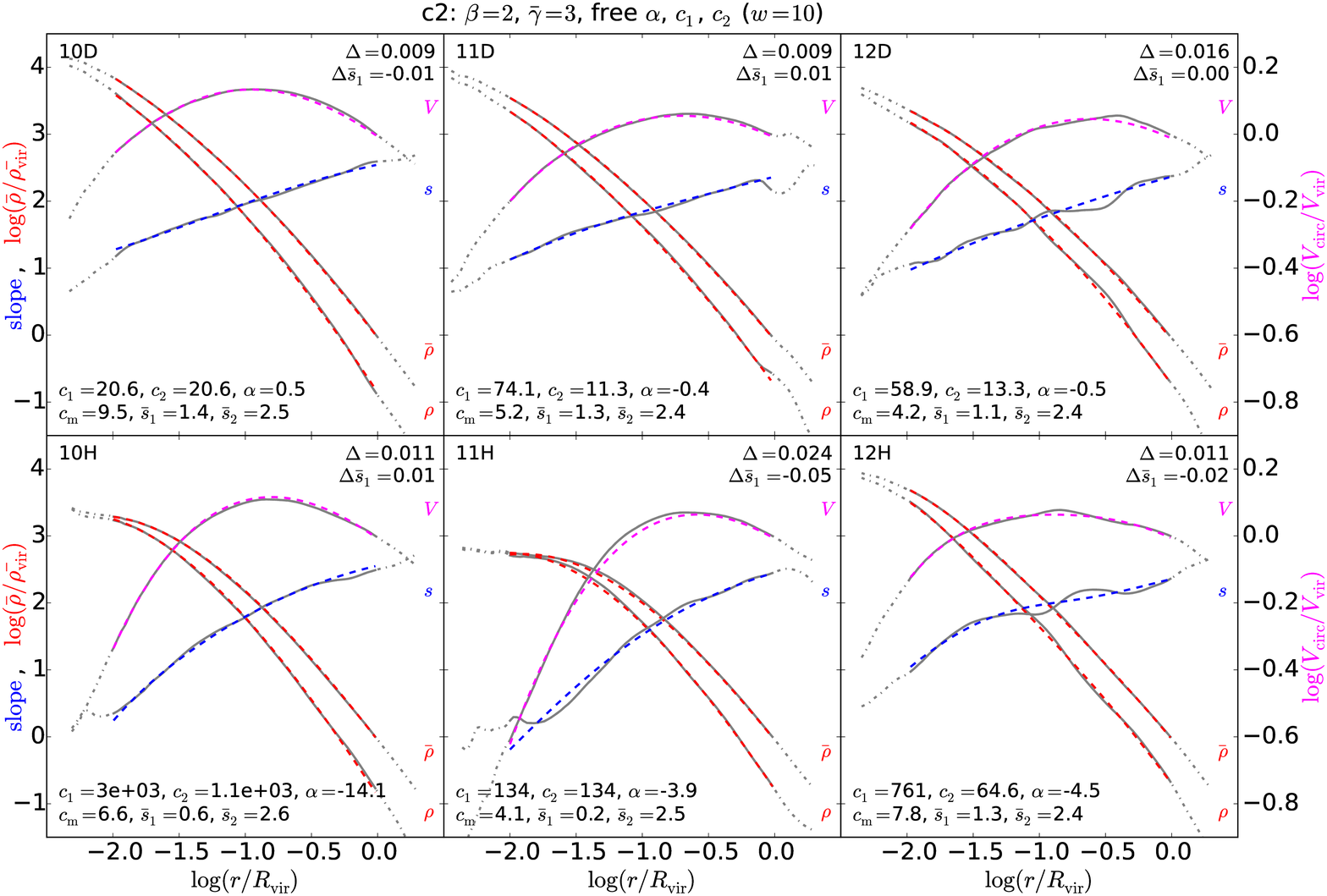}
\caption{    
Same as \fig{c2}, for the double analytic model with $\beta=2$ (model c2), 
but with enhanced weighting $w=10$ in the fit at $(0.01-0.03)\Rv$.
The fits in the inner halo are excellent, with
$\Delta \bar{s}_1=0.00-0.05$.
The overall fit is also good, with $\Delta=0.009-0.024$,
though less good than with uniform weighting.
}
\label{fig:c2_w10}
\end{figure*}

To complement the fits to simulations described in \se{fit_sims} with uniform
weighting at equally spaced log radii in the range $(0.01-1)\Rv$, 
we show here analogous fits with enhanced weight, $w=10$, in the cusp-core 
region $(0.01-0.03)\Rv$.
The results for $w=10$ are summarized in \tab{w10}, to be compared to 
\tab{w1} for uniform weighting.
\Figs{a2_w10} to (\ref{fig:c2_w10}) refer to models a2, b2, and c2,
the same functional forms as in \figs{a2} to (\ref{fig:c2}, all with
$\beta=2$,
but here with enhanced weighting in the inner halo.

\smallskip.
With the enhanced weighting in the inner halo, 
the fits in the cusp-core regions are naturally slightly better,
as measured for example by $\Delta \bar{s}_1$.
This comes at the expense of the overall quality of the fits, 
as expressed for example by $\Delta$.

\clearpage
\begin{table*}
\centering

\setlength\tabcolsep{2.8pt}

\begin{tabular}{c>{\raggedright}m{0.3cm}c>{\raggedright}m{0.3cm}cc>{\centering}p{0.3cm}cc>{\centering}p{0.3cm}cc>{\centering}p{0.3cm}cc>{\centering}p{0.3cm}cc>{\centering}p{0.3cm}cc>{\centering}p{0.3cm}cc}
\hline 
\noalign{\vskip0.05cm}
 &  & {\footnotesize{}halo \#} &  &  &  &  & \multicolumn{2}{c}{{\footnotesize{}1}} &  & \multicolumn{2}{c}{{\footnotesize{}2}} &  & \multicolumn{2}{c}{{\footnotesize{}3}} &  & \multicolumn{2}{c}{{\footnotesize{}4}} &  & \multicolumn{2}{c}{{\footnotesize{}5}} &  & \multicolumn{2}{c}{{\footnotesize{}6}}\tabularnewline[0.05cm]
\noalign{\vskip0.05cm}
 &  & {\footnotesize{}name} &  &  &  &  & \multicolumn{2}{c}{{\footnotesize{}10D}} &  & \multicolumn{2}{c}{{\footnotesize{}11D}} &  & \multicolumn{2}{c}{{\footnotesize{}12H}} &  & \multicolumn{2}{c}{{\footnotesize{}12D}} &  & \multicolumn{2}{c}{{\footnotesize{}10H}} &  & \multicolumn{2}{c}{{\footnotesize{}11H}}\tabularnewline[0.05cm]
\noalign{\vskip0.05cm}
 &  & {\footnotesize{}$\bar{s}_{1}$} &  &  &  &  & \multicolumn{2}{c}{{\footnotesize{}1.4}} &  & \multicolumn{2}{c}{{\footnotesize{}1.3}} &  & \multicolumn{2}{c}{{\footnotesize{}1.3}} &  & \multicolumn{2}{c}{{\footnotesize{}1.1}} &  & \multicolumn{2}{c}{{\footnotesize{}0.6}} &  & \multicolumn{2}{c}{{\footnotesize{}0.2}}\tabularnewline[0.05cm]
\hline 
\hline 
\noalign{\vskip0.05cm}
{\footnotesize{}\#} &  & {\footnotesize{}model} &  & \multicolumn{2}{c}{{\footnotesize{}params}} & \multicolumn{18}{c}{}\tabularnewline[0.05cm]
\cline{1-6} 
\noalign{\vskip0.1cm}
\multirow{4}{*}{{\footnotesize{}a1}} & \multirow{4}{0.3cm}{} & \multirow{4}{*}{{\footnotesize{}$\bar{\gamma}$ free, $\beta=1$}} & \multirow{4}{0.3cm}{} & {\scriptsize{}$\Delta$} & {\scriptsize{}$\Delta\bar{s}_{1}$} &  & \textbf{\scriptsize{}0.012} & \textbf{\scriptsize{}0.00} &  & \textbf{\scriptsize{}0.017} & \textbf{\scriptsize{}0.01} &  & \textbf{\scriptsize{}0.020} & \textbf{\scriptsize{}-0.02} &  & \textbf{\scriptsize{}0.024} & \textbf{\scriptsize{}0.01} &  & \textbf{\scriptsize{}0.012} & \textbf{\scriptsize{}0.01} &  & \textbf{\scriptsize{}\uline{0.013}} & \textbf{\scriptsize{}-0.02}\tabularnewline
 &  &  &  & {\scriptsize{}$c$} & {\scriptsize{}$c_{{\rm m}}$} &  & {\scriptsize{}12.7} & {\scriptsize{}9.5} &  & {\scriptsize{}18.9} & {\scriptsize{}6.4} &  & {\scriptsize{}99.6} & {\scriptsize{}10.8} &  & {\scriptsize{}16.0} & {\scriptsize{}5.1} &  & {\scriptsize{}36.4} & {\scriptsize{}7.5} &  & {\scriptsize{}17.6} & {\scriptsize{}4.6}\tabularnewline
 &  &  &  & {\scriptsize{}$\alpha$} & {\scriptsize{}$\bar{s}_{1}$} &  & {\scriptsize{}1.2} & {\scriptsize{}1.4} &  & {\scriptsize{}1.0} & {\scriptsize{}1.3} &  & {\scriptsize{}-0.1} & {\scriptsize{}1.3} &  & {\scriptsize{}0.8} & {\scriptsize{}1.1} &  & {\scriptsize{}-0.4} & {\scriptsize{}0.6} &  & {\scriptsize{}-0.5} & {\scriptsize{}0.2}\tabularnewline
 &  &  &  & {\scriptsize{}$\bar{\gamma}$} & {\scriptsize{}$\bar{s}_{2}$} &  & {\scriptsize{}2.6} & {\scriptsize{}2.5} &  & {\scriptsize{}2.3} & {\scriptsize{}2.3} &  & {\scriptsize{}2.2} & {\scriptsize{}2.2} &  & {\scriptsize{}2.4} & {\scriptsize{}2.3} &  & {\scriptsize{}2.5} & {\scriptsize{}2.4} &  & {\scriptsize{}2.6} & {\scriptsize{}2.5}\tabularnewline[0.05cm]
\hline 
\noalign{\vskip0.05cm}
\multirow{4}{*}{{\footnotesize{}a2}} & \multirow{4}{0.3cm}{} & \multirow{4}{*}{{\footnotesize{}$\bar{\gamma}$ free, $\beta=2$}} & \multirow{4}{0.3cm}{} & {\scriptsize{}$\Delta$} & {\scriptsize{}$\Delta\bar{s}_{1}$} &  & \textbf{\scriptsize{}0.007} & \textbf{\scriptsize{}-0.01} &  & \textbf{\scriptsize{}0.012} & \textbf{\scriptsize{}0.01} &  & \textbf{\scriptsize{}0.012} & \textbf{\scriptsize{}\uline{0.00}} &  & \textbf{\scriptsize{}0.018} & \textbf{\scriptsize{}0.01} &  & \textbf{\scriptsize{}\uline{0.004}} & \textbf{\scriptsize{}\uline{0.00}} &  & \textbf{\scriptsize{}0.023} & \textbf{\scriptsize{}-0.03}\tabularnewline
 &  &  &  & {\scriptsize{}$c$} & {\scriptsize{}$c_{{\rm m}}$} &  & {\scriptsize{}13.6} & {\scriptsize{}8.9} &  & {\scriptsize{}43.8} & {\scriptsize{}5.9} &  & {\scriptsize{}1.7e5} & {\scriptsize{}8.9} &  & {\scriptsize{}26.2} & {\scriptsize{}4.5} &  & {\scriptsize{}468} & {\scriptsize{}6.6} &  & {\scriptsize{}32.3} & {\scriptsize{}4.1}\tabularnewline
 &  &  &  & {\scriptsize{}$\alpha$} & {\scriptsize{}$\bar{s}_{1}$} &  & {\scriptsize{}0.6} & {\scriptsize{}1.4} &  & {\scriptsize{}0.2} & {\scriptsize{}1.3} &  & {\scriptsize{}-56.4} & {\scriptsize{}1.3} &  & {\scriptsize{}0.1} & {\scriptsize{}1.1} &  & {\scriptsize{}-5.4} & {\scriptsize{}0.6} &  & {\scriptsize{}-2.1} & {\scriptsize{}0.2}\tabularnewline
 &  &  &  & {\scriptsize{}$\bar{\gamma}$} & {\scriptsize{}$\bar{s}_{2}$} &  & {\scriptsize{}3.1} & {\scriptsize{}2.6} &  & {\scriptsize{}2.7} & {\scriptsize{}2.3} &  & {\scriptsize{}2.4} & {\scriptsize{}2.3} &  & {\scriptsize{}2.8} & {\scriptsize{}2.3} &  & {\scriptsize{}2.9} & {\scriptsize{}2.5} &  & {\scriptsize{}3.4} & {\scriptsize{}2.6}\tabularnewline[0.05cm]
\hline 
\noalign{\vskip0.05cm}
\multirow{4}{*}{{\footnotesize{}b1}} & \multirow{4}{0.3cm}{} & \multirow{4}{*}{{\footnotesize{}$\bar{\gamma}=3,$ $\beta=1$}} & \multirow{4}{0.3cm}{} & {\scriptsize{}$\Delta$} & {\scriptsize{}$\Delta\bar{s}_{1}$} &  & \textbf{\scriptsize{}0.035} & \textbf{\scriptsize{}0.03} &  & \textbf{\scriptsize{}0.054} & \textbf{\scriptsize{}0.07} &  & \textbf{\scriptsize{}0.083} & \textbf{\scriptsize{}0.10} &  & \textbf{\scriptsize{}0.056} & \textbf{\scriptsize{}0.07} &  & \textbf{\scriptsize{}0.079} & \textbf{\scriptsize{}0.08} &  & \textbf{\scriptsize{}0.062} & \textbf{\scriptsize{}\uline{0.01}}\tabularnewline
 &  &  &  & {\scriptsize{}$c$} & {\scriptsize{}$c_{{\rm m}}$} &  & {\scriptsize{}5.1} & {\scriptsize{}7.8} &  & {\scriptsize{}3.7} & {\scriptsize{}5.2} &  & {\scriptsize{}4.2} & {\scriptsize{}6.0} &  & {\scriptsize{}4.0} & {\scriptsize{}4.5} &  & {\scriptsize{}10.7} & {\scriptsize{}6.6} &  & {\scriptsize{}9.9} & {\scriptsize{}4.6}\tabularnewline
 &  &  &  & {\scriptsize{}$\alpha$} & {\scriptsize{}$\bar{s}_{1}$} &  & {\scriptsize{}1.3} & {\scriptsize{}1.4} &  & {\scriptsize{}1.2} & {\scriptsize{}1.3} &  & {\scriptsize{}1.3} & {\scriptsize{}1.4} &  & {\scriptsize{}1.1} & {\scriptsize{}1.2} &  & {\scriptsize{}0.3} & {\scriptsize{}0.7} &  & {\scriptsize{}-0.2} & {\scriptsize{}0.2}\tabularnewline
 &  &  &  &  & {\scriptsize{}$\bar{s}_{2}$} &  &  & {\scriptsize{}2.7} &  &  & {\scriptsize{}2.6} &  &  & {\scriptsize{}2.7} &  &  & {\scriptsize{}2.6} &  &  & {\scriptsize{}2.8} &  &  & {\scriptsize{}2.7}\tabularnewline[0.05cm]
\hline 
\noalign{\vskip0.05cm}
\multirow{4}{*}{{\footnotesize{}b2}} & \multirow{4}{0.3cm}{} & \multirow{4}{*}{{\footnotesize{}$\bar{\gamma}=3,$ $\beta=2$}} & \multirow{4}{0.3cm}{} & {\scriptsize{}$\Delta$} & {\scriptsize{}$\Delta\bar{s}_{1}$} &  & \textbf{\scriptsize{}0.009} & \textbf{\scriptsize{}-0.01} &  & \textbf{\scriptsize{}0.017} & \textbf{\scriptsize{}0.03} &  & \textbf{\scriptsize{}0.046} & \textbf{\scriptsize{}0.05} &  & \textbf{\scriptsize{}0.020} & \textbf{\scriptsize{}0.02} &  & \textbf{\scriptsize{}0.015} & \textbf{\scriptsize{}0.01} &  & \textbf{\scriptsize{}0.024} & \textbf{\scriptsize{}-0.05}\tabularnewline
 &  &  &  & {\scriptsize{}$c$} & {\scriptsize{}$c_{{\rm m}}$} &  & {\scriptsize{}20.6} & {\scriptsize{}9.5} &  & {\scriptsize{}9.6} & {\scriptsize{}5.5} &  & {\scriptsize{}14.2} & {\scriptsize{}7.3} &  & {\scriptsize{}12.3} & {\scriptsize{}4.5} &  & {\scriptsize{}185} & {\scriptsize{}6.6} &  & {\scriptsize{}134} & {\scriptsize{}4.1}\tabularnewline
 &  &  &  & {\scriptsize{}$\alpha$} & {\scriptsize{}$\bar{s}_{1}$} &  & {\scriptsize{}0.5} & {\scriptsize{}1.4} &  & {\scriptsize{}0.6} & {\scriptsize{}1.3} &  & {\scriptsize{}0.6} & {\scriptsize{}1.3} &  & {\scriptsize{}0.3} & {\scriptsize{}1.1} &  & {\scriptsize{}-3.4} & {\scriptsize{}0.6} &  & {\scriptsize{}-3.9} & {\scriptsize{}0.2}\tabularnewline
 &  &  &  &  & {\scriptsize{}$\bar{s}_{2}$} &  &  & {\scriptsize{}2.5} &  &  & {\scriptsize{}2.4} &  &  & {\scriptsize{}2.5} &  &  & {\scriptsize{}2.4} &  &  & {\scriptsize{}2.6} &  &  & {\scriptsize{}2.5}\tabularnewline[0.05cm]
\hline 
\noalign{\vskip0.05cm}
\multirow{4}{*}{{\footnotesize{}c1}} & \multirow{4}{0.3cm}{} & \multirow{4}{*}{{\footnotesize{}x2\enskip{} $\bar{\gamma}=3$, $\beta=1$}} & \multirow{4}{0.3cm}{} & {\scriptsize{}$\Delta$} & {\scriptsize{}$\Delta\bar{s}_{1}$} &  & \textbf{\scriptsize{}\uline{0.007}} & \textbf{\scriptsize{}\uline{0.00}} &  & \textbf{\scriptsize{}0.015} & \textbf{\scriptsize{}0.02} &  & \textbf{\scriptsize{}0.022} & \textbf{\scriptsize{}0.02} &  & \textbf{\scriptsize{}0.021} & \textbf{\scriptsize{}0.02} &  & \textbf{\scriptsize{}0.007} & \textbf{\scriptsize{}0.00} &  & \textbf{\scriptsize{}0.018} & \textbf{\scriptsize{}-0.03}\tabularnewline
 &  &  &  & {\scriptsize{}$c_{1}$} & {\scriptsize{}$c_{{\rm m}}$} &  & {\scriptsize{}13.3} & {\scriptsize{}9.5} &  & {\scriptsize{}10.1} & {\scriptsize{}6.4} &  & {\scriptsize{}14.3} & {\scriptsize{}10.8} &  & {\scriptsize{}10.3} & {\scriptsize{}4.8} &  & {\scriptsize{}27.7} & {\scriptsize{}6.6} &  & {\scriptsize{}19.3} & {\scriptsize{}4.1}\tabularnewline
 &  &  &  & {\scriptsize{}$c_{2}$} & {\scriptsize{}$\bar{s}_{1}$} &  & {\scriptsize{}2.5} & {\scriptsize{}1.4} &  & {\scriptsize{}1.3} & {\scriptsize{}1.3} &  & {\scriptsize{}1.0} & {\scriptsize{}1.3} &  & {\scriptsize{}1.6} & {\scriptsize{}1.1} &  & {\scriptsize{}5.0} & {\scriptsize{}0.6} &  & {\scriptsize{}6.0} & {\scriptsize{}0.2}\tabularnewline
 &  &  &  & {\scriptsize{}$\alpha$} & {\scriptsize{}$\bar{s}_{2}$} &  & {\scriptsize{}1.1} & {\scriptsize{}2.6} &  & {\scriptsize{}1.1} & {\scriptsize{}2.4} &  & {\scriptsize{}1.0} & {\scriptsize{}2.3} &  & {\scriptsize{}0.9} & {\scriptsize{}2.4} &  & {\scriptsize{}-0.4} & {\scriptsize{}2.6} &  & {\scriptsize{}-0.6} & {\scriptsize{}2.6}\tabularnewline[0.05cm]
\hline 
\noalign{\vskip0.05cm}
\multirow{4}{*}{{\footnotesize{}c2}} & \multirow{4}{0.3cm}{} & \multirow{4}{*}{{\footnotesize{}x2\enskip{} $\bar{\gamma}=3$, $\beta=2$}} & \multirow{4}{0.3cm}{} & {\scriptsize{}$\Delta$} & {\scriptsize{}$\Delta\bar{s}_{1}$} &  & \textbf{\scriptsize{}0.009} & \textbf{\scriptsize{}-0.01} &  & \textbf{\scriptsize{}\uline{0.009}} & \textbf{\scriptsize{}\uline{0.01}} &  & \textbf{\scriptsize{}\uline{0.011}} & \textbf{\scriptsize{}-0.02} &  & \textbf{\scriptsize{}\uline{0.016}} & \textbf{\scriptsize{}\uline{0.00}} &  & \textbf{\scriptsize{}0.011} & \textbf{\scriptsize{}0.01} &  & \textbf{\scriptsize{}0.024} & \textbf{\scriptsize{}-0.05}\tabularnewline
 &  &  &  & {\scriptsize{}$c_{1}$} & {\scriptsize{}$c_{{\rm m}}$} &  & {\scriptsize{}20.6} & {\scriptsize{}9.5} &  & {\scriptsize{}74.1} & {\scriptsize{}5.2} &  & {\scriptsize{}761} & {\scriptsize{}7.8} &  & {\scriptsize{}58.9} & {\scriptsize{}4.2} &  & {\scriptsize{}3.0e3} & {\scriptsize{}6.6} &  & {\scriptsize{}134} & {\scriptsize{}4.1}\tabularnewline
 &  &  &  & {\scriptsize{}$c_{2}$} & {\scriptsize{}$\bar{s}_{1}$} &  & {\scriptsize{}20.6} & {\scriptsize{}1.4} &  & {\scriptsize{}11.3} & {\scriptsize{}1.3} &  & {\scriptsize{}64.6} & {\scriptsize{}1.3} &  & {\scriptsize{}13.3} & {\scriptsize{}1.1} &  & {\scriptsize{}1.1e3} & {\scriptsize{}0.6} &  & {\scriptsize{}134} & {\scriptsize{}0.2}\tabularnewline
 &  &  &  & {\scriptsize{}$\alpha$} & {\scriptsize{}$\bar{s}_{2}$} &  & {\scriptsize{}0.5} & {\scriptsize{}2.5} &  & {\scriptsize{}-0.4} & {\scriptsize{}2.4} &  & {\scriptsize{}-4.5} & {\scriptsize{}2.4} &  & {\scriptsize{}-0.5} & {\scriptsize{}2.4} &  & {\scriptsize{}-14.1} & {\scriptsize{}2.6} &  & {\scriptsize{}-3.9} & {\scriptsize{}2.5}\tabularnewline[0.05cm]
\hline 
\end{tabular}

\caption{Same as \tab{w1}, but with enhanced weighting of $w=10$ in the inner
halo, $(0.01-0.03)\Rv$.
The fits are naturally better in the cusp-core region, at the expense of the
global fit, which is slightly less good.
}
\label{tab:w10}
\end{table*}

\label{lastpage}
\end{document}

\begin{figure*}
\vskip 12.0cm
\includegraphics{dekel_fig5.eps}
\caption{
Fitting the simulations with a single $\gamma=3$ profile,
with two free parameters $\alpha$ and $c$.
Equal weights are assigned at all radii, $w=1$, to be compared to
\fig{2par_w10} where $w=10$ at $(0.01=0.03)\Rv$.
The overall fit is typically slightly better here, while the fit in the inner 
halo is sometimes slightly less accurate. 
}
\label{fig:2par}
\end{figure*}

\begin{figure*}
\vskip 12.0cm
\includegraphics{dekel_fig6.eps}
\caption{
Fitting the simulations with a double $\gamma=3$ profile,
with three free parameters $\alpha$, $c_1$ and $c_2$.
Equal weights are assigned at all radii, $w=1$, to be compared to
\fig{double_w10} where $w=10$ at $(0.01=0.03)\Rv$.
The overall fit is typically slightly better here, while the fit in the inner 
halo is sometimes slightly less accurate.
}
\label{fig:double}
\end{figure*}

\begin{figure*}
\vskip 12.0cm
\includegraphics{dekel_fig7.eps}
\caption{
Fitting the simulations with three free parameters,
$\gamma$, $\alpha$ and $c$.
Equal weights are assigned at all radii, $w=1$, to be compared to
\fig{3par_w10} where $w=10$ at $(0.01=0.03)\Rv$.
The overall fit is typically slightly better here, while the fit in the inner 
\label{fig:3par}
\end{figure*}

\label{lastpage}
\end{document}